\documentclass[a4paper,longnamesfirst,useAMS,usenatbib,usedcolumn]{mn2e}

\usepackage{ifpdf}

\usepackage{amsmath}
\usepackage{amssymb}
\usepackage{times}

\usepackage{graphicx}

\usepackage{url}

\usepackage{hyperref}

\newcommand{\oiint}{\oint}

\newcommand{\myangle}{\measuredangle}

\newcommand{\sub}[1]{_{\text{#1}}}

\newcommand{\subb}[1]{_{\smash{\text{#1}}}}

\newcommand{\thi}{\vc\theta}
\newcommand{\ths}{\vc\theta\sub s}

\newcommand{\zi}{\thi}
\newcommand{\zs}{\ths}

\newcommand{\Os}{\Omega\sub s}

\newcommand{\rtext}[1]{\hspace{0.5em}\text{#1}}

\newcommand{\nablaperp}{\vc\nabla_{\!\!\perp}}
\newcommand{\nablaperpsc}{\nabla_{\!\!\perp}}
\newcommand{\nablarhat}{\vc\nabla_{\!\vc\theta}}

\newcommand{\tg}{\Delta t\sub{geom}}
\newcommand{\tp}{\Delta t\sub{pot}}

\newcommand{\sqrtsmash}[1]{\sqrt{\smash[b]{#1}}}

\newcommand{\lapl}{\vc\nabla^2}
\newcommand{\laplrhat}{\nablarhat^2}
\newcommand{\laplperp}{\nablaperp^2}
\newcommand{\lapliii}{\vc\nabla_{\!\!3}}

\newcommand{\lagrange}{{\cal L}}

\newcommand{\mr}{M\sub R}
\newcommand{\mt}{M\sub T}

\newcommand{\mat}[1]{{\mathbfss{#1}}}

\newcommand{\vc}[1]{\bmath{#1}}

\newcommand{\diff}{\text{d}}
\newcommand{\Diff}{\text{D}}

\DeclareMathOperator{\sign}{sign}

\newcommand{\deter}[1]{\det\left(#1\right)}

\newcommand{\ord}[1]{{\cal O}(#1)}
\newcommand{\ordb}[1]{{\cal O}\bigl(#1\bigr)}

\newcommand{\msigma}{\overline\sigma}
\newcommand{\mrho}{\overline\rho}

\newcommand{\psix}{\tilde\psi}

\newcommand{\sprod}{\cdot}
\newcommand{\Ds}{D\subb s}
\newcommand{\Dd}{D\subb d}
\newcommand{\Dds}{D\subb{ds}}

\newcommand{\Rm}{R_M}
\newcommand{\Psio}{\Psi\sub o}
\newcommand{\Psis}{\Psi\sub s}

\DeclareMathOperator{\heavi}{\Theta}

\newcommand{\eqlin}{\approx}

\title[From planes to spheres]{From planes to spheres: \\ About gravitational
  lens magnifications}
\author[O.~Wucknitz]
       {O.~Wucknitz$^{1,2}$\thanks{E-mail: {\tt
wucknitz@astro.uni-bonn.de}} 
  \\
$^1$ Joint Institute for VLBI in Europe, Postbus 2, 7990~AA Dwingeloo,
  The Netherlands
         \\
$^2$ Argelander-Institut f\"ur Astronomie, Universit\"at Bonn, Auf dem H\"ugel
  71, 53\,121 Bonn, Germany
}
\date{Accepted 2008 January 24. Received 2007 December 21}

\begin{document}

\maketitle

\begin{abstract}
  We discuss the classic theorem according to which a gravitational
  lens always produces at least one image with a magnification greater than
  unity. This theorem seems to contradict the conservation of total flux from
  a lensed source.  The standard solution to this paradox is based on
  the exact definition of the reference `unlensed' situation, in which the
  lens mass can either be removed or smoothly redistributed.
  
  We calculate magnifications and amplifications (in photon number and energy
  flux density) for general lensing scenarios not limited to regions close to
  the optical axis. In this way the formalism is naturally extended from
  tangential planes for the source and lensed images to complete spheres. We
  derive the lensing potential theory on the sphere and find that the Poisson
  equation is modified by an additional source term that is related to the
  mean density and to the Newtonian potential at the positions of observer and
  source.  This new term generally reduces the magnification, to below unity
  far from the optical axis, and ensures conservation of the total photon
  number received on a sphere around the source.
  
  This discussion does not affect the validity of the \emph{focusing theorem},
  in which the unlensed situation is defined to have an unchanged affine
  distance between source and observer. The focusing theorem does not
  contradict flux conservation, because the mean total magnification (or
  amplification) directly corresponds to different areas of the source (or
  observer) sphere in the lensed and unlensed situation. We argue that a
  constant affine distance does not define an \emph{astronomically} meaningful
  reference.

  By exchanging source and observer, we confirm that magnification and
  amplification differ according to Etherington's reciprocity law, so that
  surface brightness is no longer strictly conserved. At this level we also
  have to distinguish between different surface brightness definitions that
  are based on photon number, photon flux, and energy flux.
\end{abstract}


\section{Introduction}

Already \citet{einstein36} calculated the magnification of a background star
lensed by a foreground compact mass and found the classic result for the total
magnification of both images, which decreases for large impact angles (far
away from the optical axis) but always stays above unity.

At that time, gravitational lensing was regarded as a theoretical curiosity
with very little chance of leading to \emph{observable} effects. Much later,
after the first actual lens systems had been found, \citet{schneider84}
presented and proved a theorem stating that a gravitational lens with
arbitrary mass distribution will always produce at least one image with a
magnification greater than 1.  Since the theorem is thought to be valid for all
configurations of source, lens and observer, one might naively think that this
necessarily leads to an increase of the total number of photons from the
source. This would be in contradiction with the fact that lensing can neither
create nor destroy photons.  It was already pointed out by \citet{schneider84}
that such an interpretation is not correct. In order to compare the lensed
with the unlensed situation, it is essential to define the latter in a
meaningful way, which is far from trivial. In general relativity, the lens
will slightly distort the geometry of the whole Universe, so that it is
plainly impossible to add a lens without changing the configuration of source
and observer. The `reference situation' in the standard explanation is the one
in which the mass of the lens is redistributed evenly over the Universe.  It
can be shown that the comparison finds no net increase of photon number then
\citep[e.g.][]{weinberg76}. The reason for this is that two differently lensed
scenarios are compared.

Some level of uneasiness remained in the minds of several authors, because one
may think of situations in which the general validity of the theorem would
still lead to a violation of photon number conservation.  \citet{avni88}
calculated the amplification caused by a point-mass lens by dropping the
small-angle approximation of light rays with respect to the optical axis
(defined by observer and lens), which is generally used in the standard
formalism.  One might think that far from the optical axis the deflection
becomes so weak that the full calculation would add only higher-order terms
which are not relevant in the end. Still, slightly surprising, these authors
find that the amplification actually sinks below 1, which corrects for the
amplification excess close to the optical axis.  \citet{avni88} worked with a
point-mass metric in Schwarzschild coordinates, and fixed the radial
coordinate relative to the lens for the comparison of lensed and unlensed
situation.  \citet{sef} question the value of such a comparison. They argue
that there is generally no unique way to compare mean fluxes in different
space-times.  Nevertheless, approximating the calculations of \citet{avni88}
for regions close to the optical axis would directly lead to the tangential
plane approach of \citet{schneider84}. It is not entirely clear why (if at
all) the theorem loses its validity far from the optical axis.

\citet{jaroszynski96} make similar calculations, but keep the \emph{metric
  distance} between source and observer constant for the comparison. They find
that the effect of the lens changes the total area of the sphere around the
source and can thus (without contradicting photon conservation) lead to
changes in the averaged amplification. They also make a distinction between
\emph{magnification} and \emph{amplification}, which becomes relevant at the
accuracy levels required far from the optical axis.

None of these publications actually presents the magnification to second order
(in terms of the deflection potential) for arbitrary angles to the optical
axis\footnote{Close to the optical axis, the second-order terms become
  dominant for compact masses.}. They either approximate for small angles or
include only first-order terms.  One aim of our work is to extend the
calculations to include second-order terms also for large impact angles.  Even
more important is the discussion of appropriate unlensed situations to act as
a reference in order to define magnifications and amplifications.

We will also develop the formalism to treat off-axis lenses similarly to the
standard small-angle approximation and learn which part of the proof of the
magnification theorem cannot be generalised for this situation. This will be
done for arbitrary mass distributions and not only for point-mass lenses.

The outline of this paper is as follows.  We start in Sec.~\ref{sec:planar}
with a brief review of the standard planar lensing theory, which is described
in a tangential plane at the position of the lens. This includes the
magnification theorem and a simple version of its proof. In
Sec.~\ref{sec:solid} we explain our approach of discussing the apparent
magnification in terms of solid angles. In this way we can avoid the
explicit definition of an unlensed reference situation. Solid angle must
always integrate to $4\upi$, so that the magnification theorem (in these
terms) cannot hold everywhere. This discussion is equivalent to a reference
situation in which the area of the source sphere stays constant.

Section~\ref{sec:point-mass} comprises
calculations for a point-mass lens at infinity. We explicitly find that the
magnification 
sinks below 1 far from the optical axis.  General mass distributions (still
with source at infinity) are discussed in Sec.~\ref{sec:general mass sphere}.
There we develop the potential theory of lenses on the sphere as an extension
of the standard potential theory in the tangential plane. We find that the
Poisson equation is modified by an additional source term, which ensures the
validity of Gauss' theorem on the closed sphere.

The exact form of magnification matrices on the sphere for finite (and
possibly large) deflection angles is investigated in Appendix~\ref{sec:mag
  sph}. For our current discussion we only need the result that the
magnification matrix in terms of second-order derivatives of the potential
differs from the usual form only to second order in the deflection angle, so
that this difference can be neglected.

In Sec.~\ref{sec:finite ds} we extend our calculations to finite values of the
source distance and explicitly calculate deflection angles and magnifications
for point-mass lenses. By adapting the potential theory to this situation, we
find that the correction terms in Poisson equation and magnification are
directly related to the Newtonian potential at the source and observer.

The subject of reciprocity is covered in Sec.~\ref{sec:recip}, where we
exchange the role of source and observer to relate \emph{magnification} with
\emph{amplification}. These two are \emph{not} the same anymore, so that the
surface brightness is no longer conserved if we define it in a standard
physical sense. This is a direct consequence of the reciprocity theorem of
\citet{etherington33} and of the space-time modification at the position of
source and observer, caused by the gravitational field of the lens. In
anticipation of this result we try to make a clear distinction between
amplification and magnification from the beginning.

After a brief discussion of light travel times in Sec.~\ref{sec:light time} we
summarise and discuss our results in Secs.~\ref{sec:summary} and
\ref{sec:conclusions}.

\section{Planar lensing theory}
\label{sec:planar}

\subsection{Basics}
\label{sec:planar basics}

Here we only recall the most relevant aspects. The complete theory can be
found elsewhere \citep[e.g.][]{sef}.  Close to an optical axis (which can be
defined arbitrarily, conventionally by observer and lens centre), the geometry
of a gravitational lens system can be projected on the tangential plane of the
celestial sphere. This leads to the notion of the \emph{source plane} for the
true source structure and the \emph{lens} or \emph{image plane} for the
observed lensed image(s).  In these planes we define two-dimensional unit-less
vectors which are based on angular coordinates on the sphere.  We now restrict
the discussion to static, `thin' lenses, where the light is deflected only
very close to the lens plane. We can then interpret the action of the lens as
a \emph{potential time-delay}, which affects the light when it crosses the
lens plane. This direct potential delay causes a deflection and thus
changes the geometrical light path, which creates an additional
\emph{geometrical delay}.  The total delay $\Delta t$ is usually written in
terms of the Fermat potential, which comprises the geometrical and potential
parts in a normalized version:\footnote{The potential is defined modulo an
  arbitrary additive constant. We will generally neglect this constant, even
  if it diverges. Within equations, however, consistent constants are used.}
\begin{gather}
  \phi_{\zs}(\zi) = \frac{(\zi-\zs)^2}{2} - \psi(\zi)
\label{eq:fermat} \\
\Delta t = \frac{\Dd\Ds}{c\,\Dds} \phi_{\zs}
\label{eq:delta t}
\end{gather}
Here the vectors of the observed image and true source position are
$\zi=(\theta_x,\theta_y)$ and $\zs$, respectively. The distance parameters are
angular-size distances between observer and lens/source ($\Dd$, $\Ds$) and
between lens and observer ($\Dds$).  For simplicity, we assume that the lens
is embedded in a global Minkowski metric. In a cosmological situation,
Eq.~\eqref{eq:delta t} would have to be scaled by a redshift factor $1+z\sub
d$, but remains valid apart from that.

We define the apparent deflection angle $\vc\alpha$ to fulfil the lens
equation
\begin{align}
  \zs = \zi - \vc\alpha(\zi)
\label{eq:lenseq}
\rtext{.}
\end{align}
Deflection and time-delay are related by Fermat's theorem: Images form at
stationary points of the time-delay function. These can be minima, maxima or
saddle-points. Critical images with degenerate Hessian matrix (see below,
e.g.\ Einstein rings of point sources) are unstable and shall not be discussed
here.  We thus take the derivative of the Fermat potential in
Eq.~\eqref{eq:fermat} and substitute the lens equation \eqref{eq:lenseq} to
find that the deflection angle is the gradient of the potential,
\begin{align}
  \vc\alpha(\zi) &= \vc\nabla \psi (\zi) \rtext{.}
\end{align}
If we calculate the potential from the mass distribution, we can prove that it
obeys a Poisson equation
\begin{align}
  \lapl \psi(\zi) &= 2\kappa(\zi) \rtext{.}
\end{align}
The convergence $\kappa$ is the projected surface mass density $\Sigma$ in
units of the critical density
\begin{align}
\Sigma\sub c &= \frac{c^2}{4\upi G}\frac{\Ds}{\Dd\Dds}
\label{eq:sigmac}
\rtext{.}
\end{align}
In more detail, we can express the Hessian matrix,
consisting of the second-order derivatives of $\psi$, in terms of the
convergence and the two-dimensional external shear
$\vc\gamma=(\gamma_x,\gamma_y)$:
\begin{align}
  \mat H =
\begin{pmatrix}
  \psi_{xx} & \psi_{xy} \\
  \psi_{yx} & \psi_{yy}
\end{pmatrix}
\label{eq:hesse}
= \begin{pmatrix}
  \kappa+\gamma_x & \gamma_y \\
  \gamma_y & \kappa-\gamma_x
\end{pmatrix}
\end{align}

\subsection{The magnification theorem}
\label{sec:theorem}

This theorem was presented by \citet{schneider84} and is described in a
broader context by \citet{sef}. Here we only want to make a plausibility
proof, without claiming to be fully mathematically rigorous.  The local
magnification matrix $\mat M$ is the inverse Jacobian of the lens mapping,
Eq.~\eqref{eq:lenseq}:
\begin{align}
  \mat M^{-1} = \frac{\upartial \zs}{\upartial \zi} = \mat 1-\mat H
\label{eq:magn eq}
\end{align}
The scalar magnification $\mu$ is the determinant of this matrix,
\begin{align}
  \mu^{-1} = \deter{\mat 1 - \mat H} = (1-\kappa)^2-\gamma^2
\label{eq:mag 1}
\rtext{,}
\end{align}
with $\gamma=|\vc\gamma|$. The sign of $\mu$ defines the parity of the image.

For large $\zi$, the Fermat potential will be dominated by the first term in
Eq.~\eqref{eq:fermat} for all realistic mass distributions, i.e.\ it grows
quadratically with $\zi$. Local mass concentrations will produce minima in
$\psi$ that can diverge for point masses, but can not lead to positive
discontinuous peaks in $\psi$. It is thus plausible that at least one local
minimum of $\phi$ exists. This point is a solution of the lens equation and
corresponds to a real image. Being a minimum, the Hessian of the Fermat
potential, $\mat 1-\mat H$, must be positive definite.  The eigenvalues can be
easily derived to be $1-\kappa\pm\gamma$, which must thus both be positive.
If we add the condition that the convergence or density $\kappa$ is
non-negative, we find $0<\mu^{-1}<1$. In other words, this image has positive
parity and a magnification $\mu>1$. As a result, the total magnification of
all images corresponding to one source position is always greater than unity.

The fact that lensing increases the observed flux density in all
directions\footnote{For the moment we assume the equivalence of magnification
  and amplification, which can be proven in this framework.}, no matter where
the lens is positioned, raises the question how this can be consistent with
conservation of total flux.  This issue is discussed by \citet{schneider84},
\citet{sef} and references therein. The basic argument is that one has to
thoroughly define the unlensed situation, relative to which the magnifications
are calculated. Adding a lens modifies the geometry of the Universe, so that
the comparison is not trivial.  This can be illustrated by a thought
experiment in which we shrink the size of the Universe by some factor. This
would move \emph{all} sources closer to \emph{all} observers, so that they
would all see an increased flux density. Still, there is of course no
contradiction with energy or flux conservation at all.

An important example for total net magnifications is the case of a point-mass
lens, where the calculations were presented long before the general theorem
\citep{einstein36,refsdal64a}.  The total magnification of the two
images\footnote{A real Schwarzschild lens would additionally produce an
  infinite number of images very close to the lens
  \citep[e.g.][]{virbhadra00}. Something similar, although not quantitatively
  the same, happens with the deflection angle according to Eq.~\eqref{eq:defl
    sphere m}. These images are extremely demagnified and can be safely
  neglected here.} as a function of the angular separation $\theta\sub s$
between lens and source is given by the standard expression
\begin{align}
  \mu\sub{tot} = \frac{\theta\sub s^2+2m}{\theta\sub s \sqrt{\theta\sub s^2+4m}} \rtext{,}
\end{align}
which is greater than unity everywhere.  Here we have defined
\begin{align}
  m=\frac{4GM}{c^2}\frac{\Dds}{\Dd\Ds}\quad\to\quad\frac{4GM}{c^2\, \Dd}
  \quad\text{(for $\Ds\gg\Dd, \Dds\approx\Ds$),}
\label{eq:m def}
\end{align}
which is the square of the Einstein radius.  We can calculate the excess
magnification by integrating over the tangential plane:
\begin{align}
  \Delta\mu &= \iint\diff^2 \theta\sub{s} \, (\mu\sub{tot}-1) = 2\upi\, m
\label{eq:excess mag}
\end{align}
The order of magnitude of this excess is plausible, because the lens produces
significant additional magnifications (of the order 1) over a region
corresponding to the Einstein ring with radius $\sqrt{m}$.

We will later find that the solution of the magnification paradox has to do
with terms of $\mu^{-1}$ that are linear in the deflection. This can be
anticipated by the fact that the excess in Eq.~\eqref{eq:excess mag} is also
linear in $m$. Counting only the terms linear in the deflection (linear in
$\psi$, $\kappa$, $\gamma$) of Eq.~\eqref{eq:mag 1}, we find
\begin{align}
  \mu^{-1} \eqlin 1-H_{xx}-H_{yy} = 1- \lapl\psi = 1-2\kappa \rtext{.}
\label{eq:mu lin poisson}
\end{align}
The first-order magnification can only be reduced below 1 if either $\kappa$
is negative or the Poisson equation is modified in an appropriate way.  We do
expect the latter to happen when we leave the tangential plane and consider
the sphere. The divergence (source density) of the deflection field
$\vc\nabla\!\sprod \vc\alpha=\lapl\psi$ cannot be positive everywhere,
because of Gauss' law. There simply is no chance for the flux of $\vc\alpha$
to escape from the closed sphere, so that the integrated divergence must
vanish.

\section{Solid angles}
\label{sec:solid}

To avoid most of the complications of arbitrary definitions for reference
situations, we want to interpret the apparent paradox in a different way.
Instead of discussing \emph{fluxes}, we want to investigate how \emph{solid
  angles} are mapped from the observed image sphere to the true source sphere,
which we put at an infinite distance, where the metric perturbation by the
lens can be neglected.
Using solid angles instead of areas on the source sphere, we can avoid
a possible scaling of the total area. The total solid angle always remains
$4\upi$. 

We do not restrict the discussion to small regions around the optical axis
(corresponding to the tangential plane) but instead allow for arbitrary impact
angles.

\subsection{Conservation of solid angle}
\label{sec:cons solid angle}

We start by integrating the total apparent solid angle on the image sphere and
transform this into an integral over the source sphere. For this we assume
that all lines of sight end somewhere on the source sphere and that every part
of the source sphere can be seen in at least one direction, i.e.\ we neglect
possible horizons which would add higher order corrections.
\begin{gather}
  4\upi = \oiint\diff\Omega(\vc\theta)
  = \oiint\diff\Os(\vc\theta\sub s)\, \mu\sub{tot} (\vc\theta\sub s) \\
  \mu\sub{tot}(\vc\theta\sub s) = \sum_{\vc\theta(\vc\theta\sub s)} \bigl|
  \mu(\vc\theta) \bigr| = \sum_{\vc\theta(\vc\theta\sub s)} \left|
    \frac{\diff\Omega}{\diff\Os}(\vc\theta) \right|
\label{eq:mag solid}
\end{gather}
The sum is taken over all images corresponding to the respective source
position. We learn that the total magnification, defined in terms of solid
angle elements, averaged over the source sphere \emph{must} be equal to unity.
Even if it is valid close the the optical axis, the magnification theorem
cannot hold in other regions where the excess has to be compensated.

We conclude that if the magnification theorem would hold in this situation, we
would have a real paradox that had to be taken seriously.  We will learn later
that the magnification theorem \emph{does} indeed become invalid for large
impact angles. In these regions the primary image is only very weakly
deflected. Any additional images will be strongly deflected (invalidating the
weak-deflection approximation), but highly demagnified. In the following
discussion we will mostly neglect these images, which allows us to consider
the magnification as a function of image positions instead of the total
magnification as a function of source position.  Fig.~\ref{fig:mu pm}
shows that this simplification is justified.

\begin{figure}
  \centering
  \includegraphics[width=0.3\textwidth]{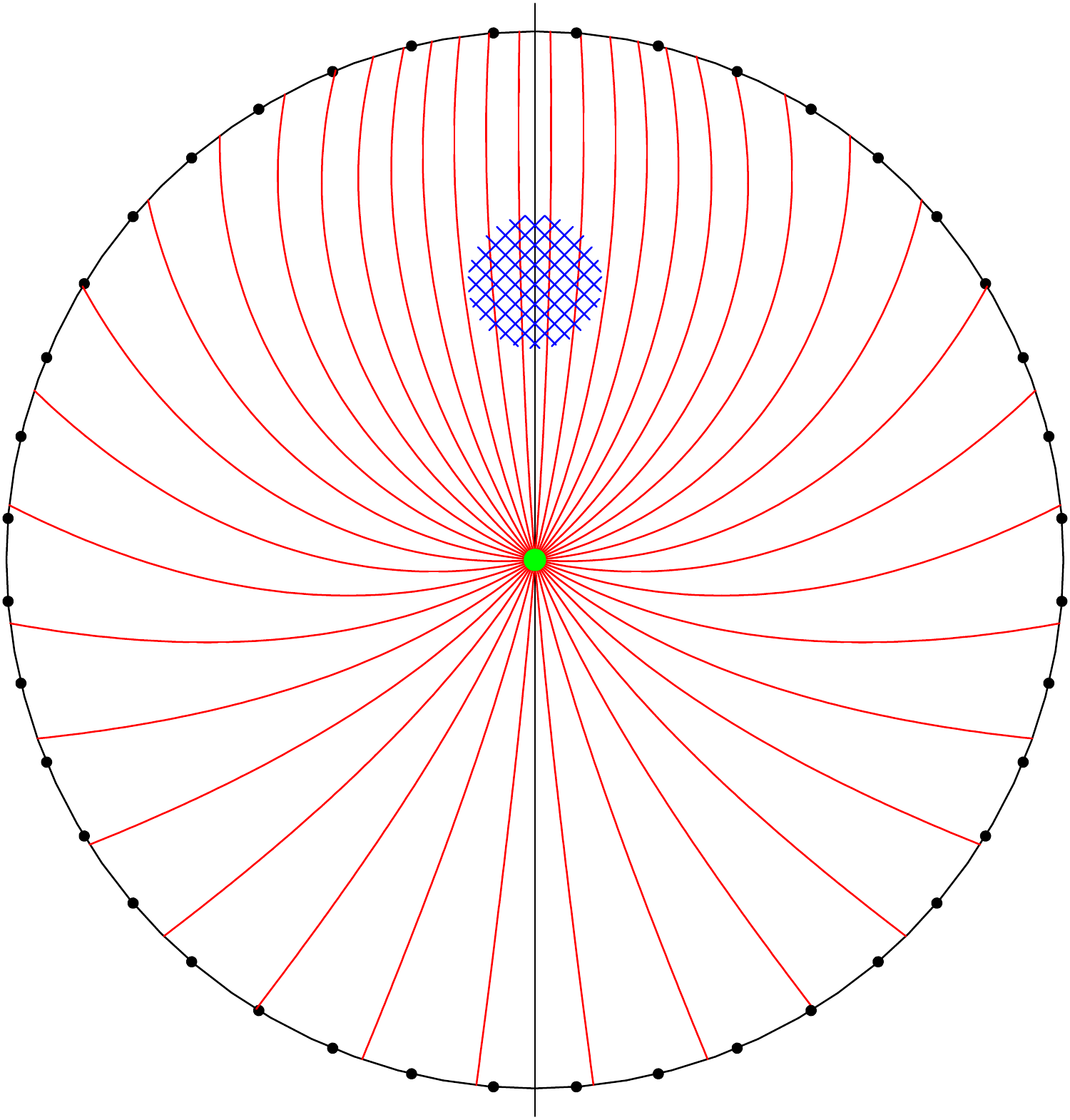}
\caption{Deflected lines of sight with the observer in the centre and the
  source sphere as circle. The vertical line defines the optical axis, the
  hatched circle shows the lens. The regularly spaced dots on the source
  sphere label the corresponding image positions (straight projections of the
  lines of sight).  }
\label{fig:defl}
\end{figure}

Figure~\ref{fig:defl} illustrates why the magnification theorem can be valid
around the optical axis but not everywhere. We see that the lens increases the
density of lines of sight on the source sphere around the (forward) optical
axis, corresponding to a magnification greater than unity.  On the tangential
plane, this would be possible everywhere, because lines of sight (or solid
angle on the sky) can be `borrowed from infinity'. In contrast to this, the
total solid angle must be conserved on the compact sphere. This necessarily
leads to magnifications \emph{less} than unity somewhere, seen in the lower
parts of the diagram.  Note that, for simplicity, a lens without multiple
imaging is shown here. The lens mass distribution was chosen in such a way
that the magnifications can be seen already in this two-dimensional
cross-sectional view. Generally the density of lines of sight only shows in
the full three-dimensional geometry.

\subsection{Reference situations and the refraction analogue}

We now know that for solid angles, measured for sources at infinity, the
magnification theorem cannot hold. However, we have to be very specific about
the situations that are compared to define a magnification.
The magnification according to Eq.~\eqref{eq:mag solid} defines the ratio of
solid angles of a cone of light rays measured by the observer and at infinity.
In astronomy we are interested in the ratio of solid angles of one and the
same source seen with and without the action of the lens.
Naively, one would define the unlensed reference situation in such a way
that the distance between observer and source is kept constant, but distances
are not uniquely defined in a curved space-time so that this approach is not
unique.
The picture of solid angles and a very large source sphere can also be
interpreted as a reference situation in which the total area of the source
sphere is the same in the lensed and unlensed situation.
Flux conservation then demands a mean magnification of unity, so that
general validity of the magnification theorem cannot be expected.

We want to postpone the discussion of an appropriate reference
situation by redefining the problem. It is well-known that the gravitational
light deflection can be described as the action of a refractive medium that
fills an Euclidean space, whose coordinates are identified with those used in
the standard form of the weak-field metric equation \eqref{eq:weak field
metric}. In this model, the unlensed situation would obviously be the one
with all distances unchanged but the refractive medium removed. Solid angles
can then directly be identified with areas on the source sphere, so that the
reasoning about solid angles in Sec.~\ref{sec:cons solid angle} inevitably
leads to a failure of the magnification theorem \emph{in this situation.}
Since this contradicts the result of Sec.~\ref{sec:planar}, even though the
formalism is exactly the same, we have to identify and abandon the
inadmissible assumptions or approximations. Once the formalism has been
generalised, we can come back to the definition of reference situations.

As an alternative to refraction we can refer to the Newtonian picture of the
deflection of massive particles moving with the speed of light. Modulo a
factor of 2, the deflection follows exactly the same law as in general
relativity, but the geometry of space and time are not altered.

\section{Sources at infinite distances}
\label{sec:infinite Ds}

\subsection{Point-mass lens}
\label{sec:point-mass}

We know that close to the optical axis the magnification theorem holds. From
the discussion of Fig.~\ref{fig:defl}, we expect that it breaks down at larger
impact angles. Therefore we have to generalize the usual formalism to allow
source positions far from the lens centre.  However, we will keep the
weak-field approximation so that deflection angles still have to be small.

We describe the lens(es) as small perturbation of an asymptotic Minkowski
metric instead of a more realistic Robertson-Walker metric.  This simplifies
the treatment without obscuring the apparent magnification paradox or its
solution.  We use the static weak-field metric in isotropic coordinates, so
that locally observed angles (and solid angles) can be derived directly from
the spatial coordinates $\vc r=(x,y,z)$.  In terms of the Newtonian potential
$\Psi$, the line element can be written as
\begin{align}
  \diff s^2 = \left( 1+\frac{2\Psi}{c^2} \right) c^2 \diff t^2 - \left(
    1-\frac{2\Psi}{c^2} \right) \diff \vc r^2
\label{eq:weak field metric}
\rtext{.}
\end{align}
This leads to the geodesic equation for light in first order approximation
\begin{align}
  \ddot{\vc r} = -2\frac{\dot{\vc r}^2}{c^2} \vc\nablaperp \Psi \rtext{,}
\label{eq:geodesic}
\end{align}
where $\vc\nablaperp$ denotes the transversal part (relative to $\dot{\vc
  r}$) of the gradient, and dots stand for derivatives with respect to the
affine parameter.  Within first order (in $\Psi$), the geometrical length or
local time can also be used equivalently.
 
\subsubsection{Deflection angle}
\label{sec:deflection pm}

\begin{figure}
  \centering
  \includegraphics[width=0.475\textwidth]{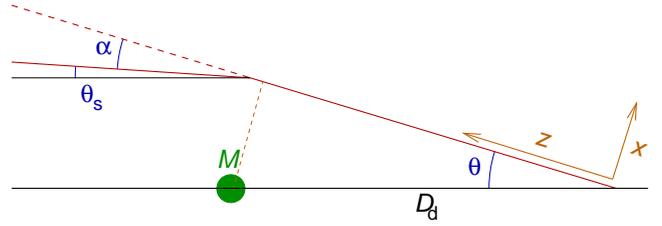}
\caption{Geometry of the light deflection with a source at infinity. The
  angles $\theta$, $\theta\sub s$ and $\alpha$ are positive in our sign
  convention.
Here and in all following geometrical considerations, we pretend that space is
Euclidean (and space-time Minkowski) and treat the light deflection as a
non-geometrical effect. The figures represent the geometry of the coordinates
used to define the metric in Eq.~\eqref{eq:weak field metric}. Isotropy of the
metric ensures that all angles correspond to real locally observable
angles. Straight lines in the figures are not necessarily geodesics in real
space or space-time. 
}
\label{fig:geom}
\end{figure}

We want to calculate the deflection angle of a point mass $M$ with a source at
infinity.  We restrict ourselves to the plane defined by lens, source and
observer, which contains the complete light ray. The coordinates have their
origin at the observer, $z$ is measured along the unperturbed line of sight,
$x$ in the perpendicular direction (Fig.~\ref{fig:geom}).  The deflection
angle is defined according to the standard form of the lens equation
\eqref{eq:lenseq}.
The lens, located at
\begin{align}
  x_0=-\Dd\sin\theta\rtext{,} \quad z_0=\Dd\cos\theta\rtext{,}
\label{eq:x0z0}
\end{align}
produces a Newtonian potential of
\begin{align}
  \Psi(x,z) = -\frac{GM}{\sqrt{(x-x_0)^2+(z-z_0)^2}} \rtext{.}
\end{align}
The deflection angle is calculated by integrating Eq.~\eqref{eq:geodesic}
along the unperturbed line of sight. It is defined to be positive for
deflection in the negative $z$-direction. We use $z$ as affine parameter:
\begin{align}
  \alpha &= - \Delta \dot x = - \int_0^\infty \diff z\, \ddot{x} (x=0,z)
\label{eq:alpha first}
\\
&= \frac{2}{c^2} \int_0^\infty \diff z\, \nablaperpsc\Psi (x=0,z)
\label{eq:alpha int Psi}
\\
&= -\frac{2GM}{c^2}\, x_0 \int_0^\infty \diff z\, \Bigl[ x_0^2 + (z-z_0)^2
\Bigr]^{-3/2}
\\
&= -\frac{2GM}{c^2} \frac{1+z_0 \bigl(x_0^2+z_0^2\bigr)^{-1/2}}{x_0}
\\
&= \frac{m}{2} \frac{1+\cos\theta}{\Dd\,\sin\theta} = \frac{m}{2}
\cot\frac{\theta}{2}
\label{eq:defl sphere m}
\end{align}
In the limit of small $\theta$, i.e.\ close to the optical axis, we recover
the standard result $\alpha=m/\theta$, with the definition of $m$ from
Eq.~\eqref{eq:m def}.

Note that for large $\theta$ the deflection is not confined to regions small
compared to $\Dd$, but takes place over a range $\sim |x_0|$. This means that
not even the point-mass lens can be considered as \emph{thin} anymore.

\subsubsection{Potential}

The deflection angle in Eq.~\eqref{eq:defl sphere m} can be written as the
derivative of a potential $\psi$, which is defined on the sphere as
\begin{align}
  \psi\sub{pm}(\theta) = m\ln\sin\frac{\theta}{2} = \frac{m}{2} \ln
  (1-\cos\theta) + \text{const} \rtext{.}
\label{eq:psi pm infty}
\end{align}

\begin{figure*}
  \centering
\ifpdf%
\includegraphics[width=0.47\textwidth]{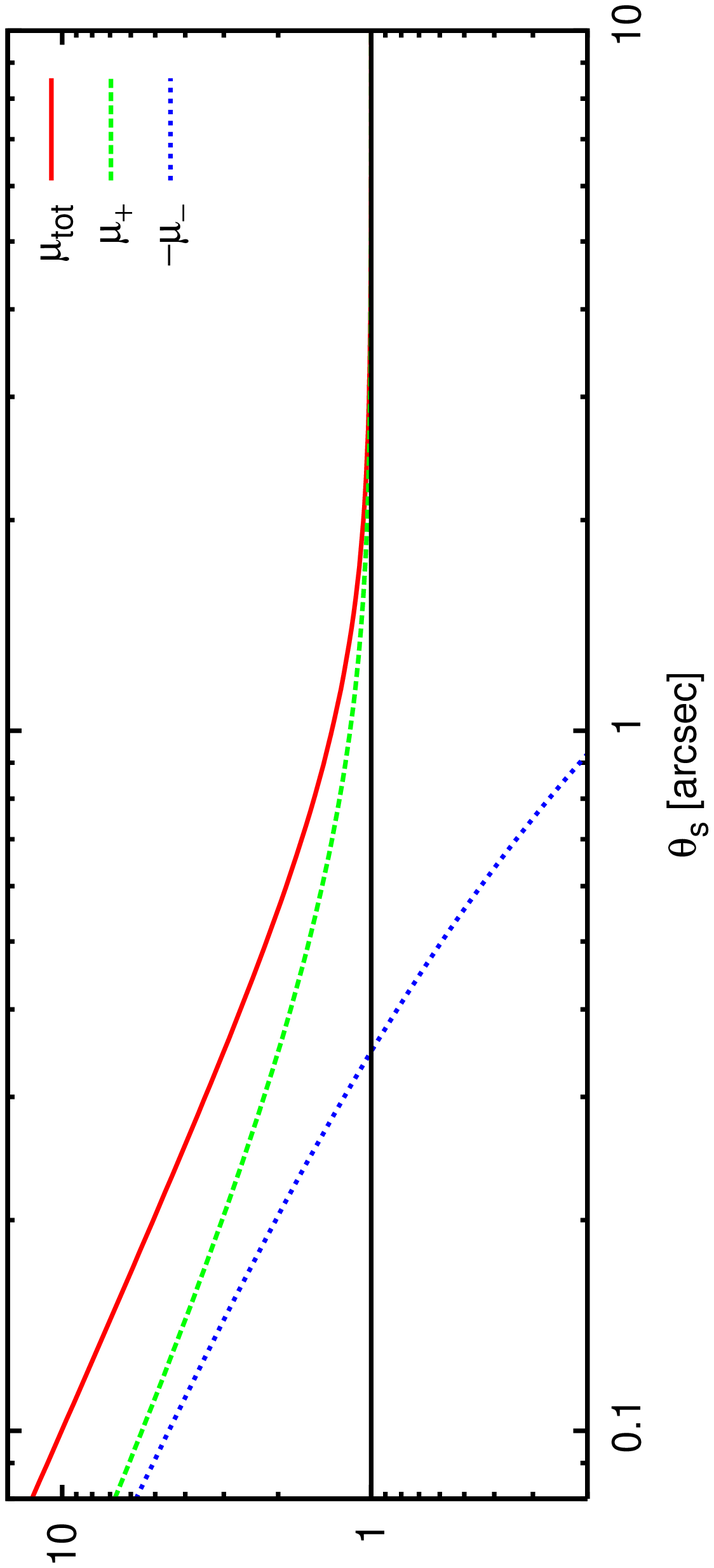}%
\else%
\includegraphics[angle=-90,width=0.47\textwidth]{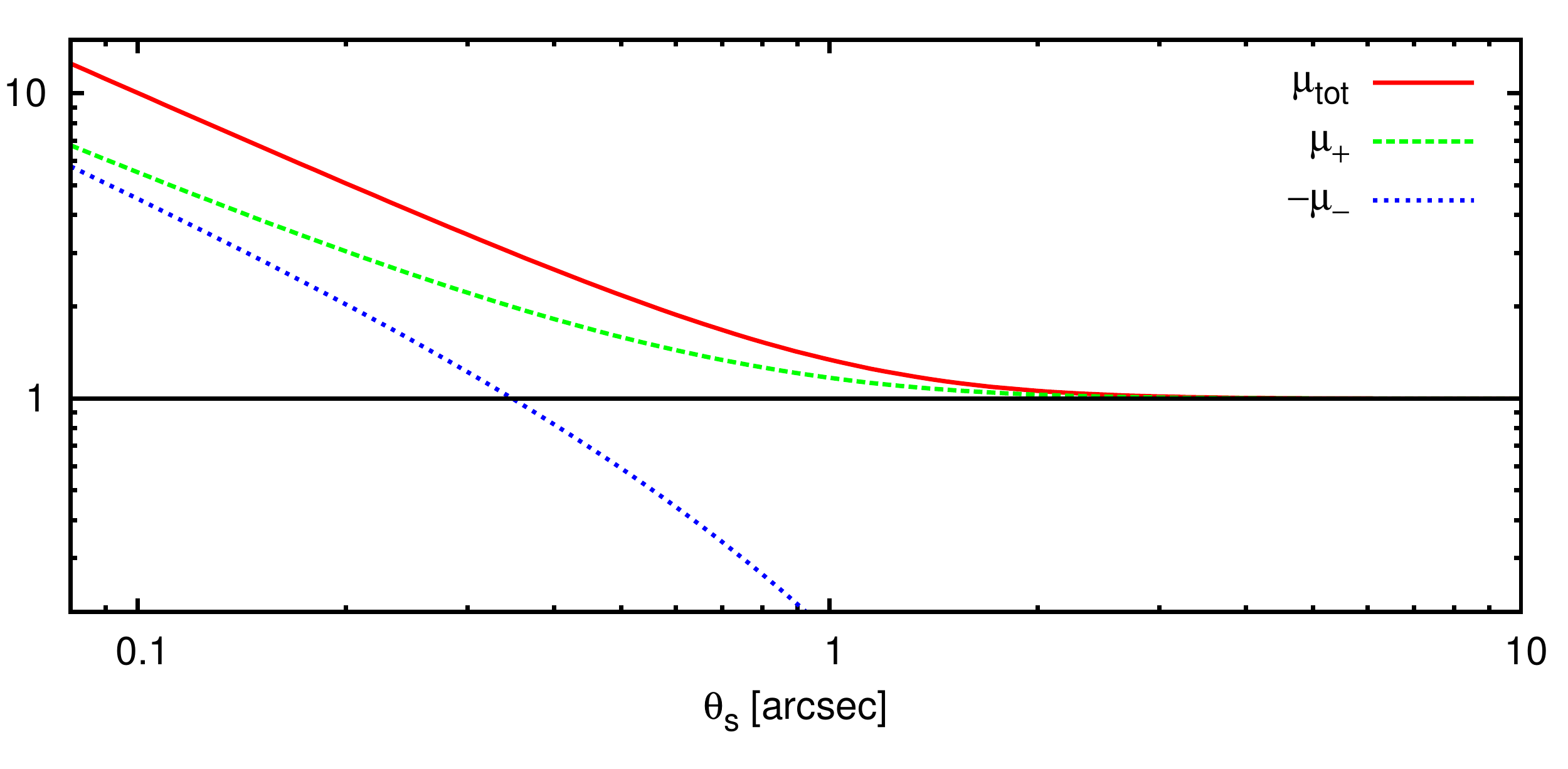}%
\fi%
\hspace{0.05\textwidth}%
\ifpdf%
\includegraphics[width=0.47\textwidth]{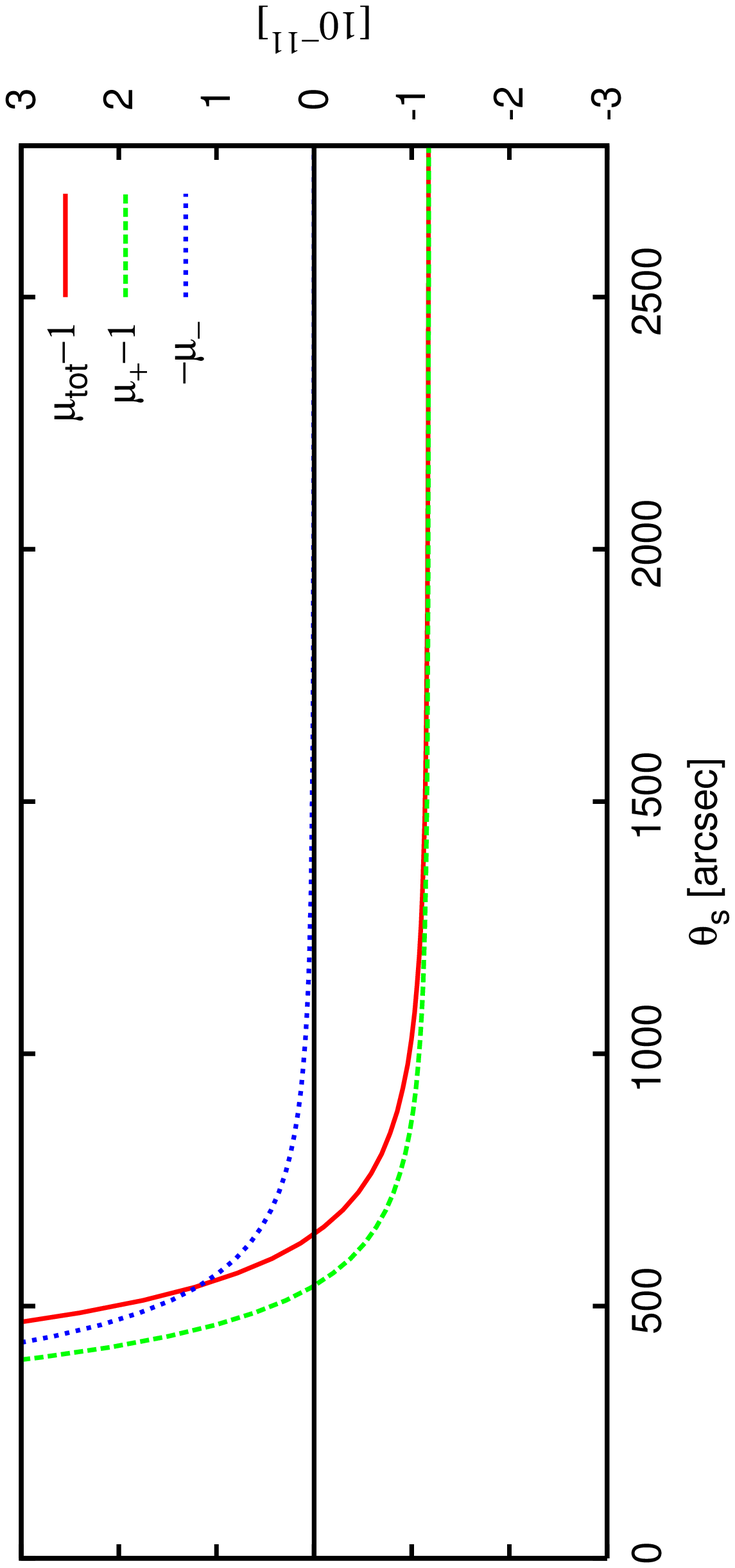}%
\else%
\includegraphics[angle=-90,width=0.47\textwidth]{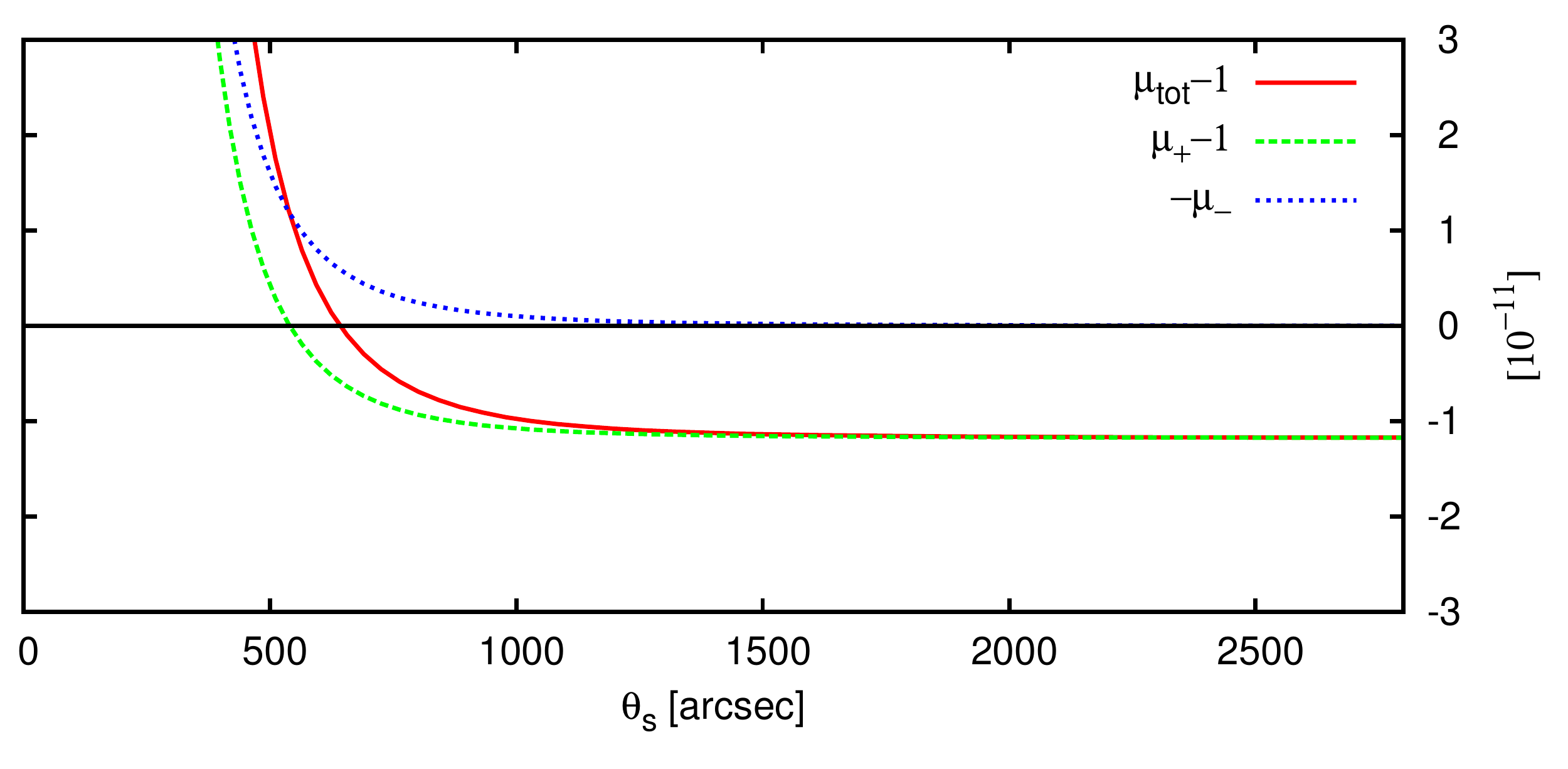}%
\fi%
\caption{Magnification of both images (positive and negative parity) and the
  total magnification as a function of the source position for a lens of
  $m=(1'')^2=2.35\cdot10^{-11}$. For small angles (left), the magnification is
  indistinguishable from the classic result. Far from the optical axis
  (right), the magnification approaches $1-m/2$ instead of 1. The
  magnification theorem does not hold in this region.  The total magnification
  and the magnification of only the positive parity image are
  indistinguishable for very large impact angles. This justifies our approach
  of not discussing the total magnification of all images but only that of the
  primary image ($\mu_+$).  }
\label{fig:mu pm}
\end{figure*}

\subsubsection{Magnification}

Consider as source a thin concentric annulus around the lens, as seen by the
observer. With radius $\theta\sub s$ and width $\diff\theta_s$, the solid
angle is $\diff\Os=2\upi\sin\theta\sub s\,\diff\theta\sub s$. This annulus
will be observed at radius $\theta$ with width $\diff\theta$ and solid angle
$\diff\Omega=2\upi\sin\theta\,\diff\theta$.  We separate the magnification
into the tangential and radial parts, $\mt$ and $\mr$, respectively:
\begin{align}
  \mu^{-1} &= \frac{\diff\Os}{\diff\Omega} = \mt^{-1} \mr^{-1}
\label{eq:mag tot tang rad}
\\
\mt^{-1} &= \frac{\sin\theta\sub s}{\sin\theta} =
\frac{\sin(\theta-\alpha)}{\sin\theta} = \cos\alpha -\cot\theta \sin\alpha
\label{eq:mag tang}
\\
\mr^{-1} &= \frac{\diff\theta\sub s}{\diff\theta} = 1-
\frac{\diff\alpha}{\diff\theta}
\label{eq:mag rad}
\\
&= 1+\frac{m}{4}\sin^{-2}\frac{\theta}{2}
\label{eq:mag rad2}
\end{align}
The magnification we get from Eqs.~(\ref{eq:mag tot tang rad}--\ref{eq:mag
  rad}) with the deflection angle from Eq.~\eqref{eq:defl sphere m} is plotted
in Fig.~\ref{fig:mu pm}.  The deflection angle and its derivative are correct
only to first order in $m$, so that the first order is also sufficient for the
radial and tangential magnifications\footnote{This is not true for the scalar
  magnification $\mu$ (the determinant of the magnification matrix), where the
  second order terms, resulting from combinations of first order terms in
  $\mt$ and $\mr$, are responsible for the magnifications in the classical
  lensing regime close to the optical axis.
  It is ironic that our generalisation of the standard formalism introduces
  additional terms of first order to $\mu$, whilst the dominating classical
  part is of second order.}.  Up to this order, we find
\begin{align}
  \mt^{-1} &= 1 - \alpha \cot\theta \rtext{,}
\label{eq:mt lin}
\\
\mu^{-1} &= 1 + \frac{m}{2} - \frac{m^2}{4}
\frac{\cos\theta}{(1-\cos\theta)^2}
\label{eq:m m2 new}
\rtext{.}
\end{align}
For large $\theta$, the $m/2$ term in Eq.~\eqref{eq:m m2 new} is dominant,
whilst in the normal lensing regime, the $m^2$ term becomes more important.
Expanded in powers of $\theta$, we find
\begin{align}
  \mu^{-1} = 1+\frac{m}{2} - \frac{m^2}{\theta^4} \Bigl[1 + \ordb{\theta^2}
  \Bigr] \rtext{.}
\end{align}
The corrections in the small-angle regime (last term) are not relevant in our
context. The isotropic correction $m/2$, on the other hand, leads to
violations of the magnification theorem. Far away from the optical axis, the
magnification sinks below unity. The transition region with a magnification of
1 is around $\theta=\sqrt[4]{2m}$, approximately the square root of the
Einstein radius. For typical galaxy lensing cases, this is a few arc-minutes
from the lens, far away from the strong-lensing regime but still at a small
angle. For microlensing, the transition region is even closer.

The small correction of the order $m$ ($10^{-10}$ in typical galaxy scale
lenses) is completely irrelevant for all practical calculations. Nevertheless,
this term is responsible for the conservation of total flux or solid angle.
Integrated over the complete celestial sphere, it leads to a \emph{deficit} of
$2\upi\, m+\ord{m^2}$, which exactly compensates for the \emph{excess} found
in Eq.~\eqref{eq:excess mag}.
Because the modification is of first order in $m$, the following discussion
will concentrate on first-order effects.

\subsection{General mass distributions on the sphere}
\label{sec:general mass sphere}

\subsubsection{Potential}

Knowing the potential for a point-mass lens from Eq.~\eqref{eq:psi pm infty},
we can easily generalize for arbitrary mass distributions, defined by the
three-dimensional mass density $\rho(\vc r)$:
\begin{align}
  \psi(\vc\theta) &= \frac{4G}{c^2} \iiint\diff^3 r'\, \frac{\rho(\vc
    r')}{r'}\ln\sin\frac{\myangle(\vc\theta,\vc r')}{2}
\label{eq:phi full 1}
\\
&= \frac{2G}{c^2} \iiint\diff^3 r'\, \frac{\rho(\vc r')}{r'}\ln
\left(1-\frac{\vc\theta\sprod \vc r'}{r'}\right) + \text{const}
\label{eq:phi full 2}
\end{align}
The direction is denoted by the unit vector $\vc\theta$ with $\vc r=\vc\theta
r$.  We separate the radial integration from the tangential part on the sphere
by introducing the normalized surface mass density\footnote{Note that in this
  radial projection the influence of mass elements at distances $r$ does not
  scale with $r$ but with $1/r$, see Eqs.~(\ref{eq:phi full 1}--\ref{eq:phi
    full 2}).
  
  For a thin shell around the observer at a certain distance $\Dd$, we find $
  \sigma(\vc\theta) =(4\upi G/c^2) \Dd \int \diff r\, \rho(\vc
  r)=\Sigma/\Sigma\sub c$ as in the standard formalism. Compare with
  Eq.~\eqref{eq:sigmac}. }
\begin{align}
  \sigma(\vc\theta) = \frac{4\upi G}{c^2} \int_0^\infty \diff r\, r\,\rho(\vc
  r)
\label{eq:sigma def1}
\end{align}
to find
\begin{align}
  \psi(\vc\theta) = \frac{1}{2\upi} \oiint \diff\Omega'(\vc\theta') \,
  \sigma(\vc\theta')\, \ln\left( 1-\vc\theta\sprod \vc\theta' \right)
  +\text{const} \rtext{.}
\end{align}
In the tangential plane, this corresponds to the integral
\begin{align}
  \psi(\vc\theta) = \frac{1}{\upi} \iint \diff^2\theta' \,
  \sigma(\vc\theta')\, \ln|\vc\theta-\vc\theta'| +\text{const} \rtext{.}
\end{align}
Even though this formalism is valid for arbitrary mass distributions, it is
correct only to first order in $\rho$. This means it cannot be used for
multi-plane strong lensing, where the change of impact parameter in one lens
plane as a result of the deflection in another plane becomes relevant.

\subsubsection{Poisson equation}
\label{eq:poisson infty}

The Poisson equation for the spherical potential $\psi$ can be derived with
very little formal calculations. We determine the flux $F$ of the deflection
field $\vc\alpha$ through a circle of radius $\theta$ around the point mass
$m$, corresponding to $\sigma(\vc\theta)=\upi\, m\,\delta^2(\vc\theta)$.
This is the product of $\alpha$ with the circumference of the circle,
\begin{align}
  F(\theta)=\alpha(\theta) \; 2\upi\sin\theta = \upi m (1+\cos\theta)
  \rtext{.}
\end{align}
In the limit of $\theta\to0$, this starts with $F(0+)=2\upi m$ and then
decreases linearly with the enclosed area $A$ of the circle:
\begin{align}
  F(\theta) &= 2\upi m -\frac{m}{2}A(\theta) & A(\theta)&=2\upi (1-\cos\theta)
\end{align}
We know from Gauss' theorem that this flux equals the integrated divergence of
the field ($=$ Laplacian of the potential), so that we can write the Poisson
equation as
\begin{align}
  \lapl \psi\sub{pm}(\vc\theta) = 2\upi\, m\,\delta^2 (\vc\theta)
  -\frac{m}{2} \rtext{.}
\label{eq:poisson sphere pm}
\end{align}
The $\delta$ term is the same as on the tangential plane. The mass acts in the
usual way as source for the deflection field. However, in contrast to the
tangential plane, the sphere is closed, so that sources and sinks of the field
must compensate each other; the field lines cannot extend to infinity.
Without the second term in Eq.~\eqref{eq:poisson sphere pm}, the field lines
would continue and meet at $\theta=\upi$ where they would form an additional
singularity of mass $-m$. Instead, the field `decays' to ensure a vanishing
total integrated $\lapl\psi$.

For an arbitrary mass distribution $\sigma$, the Poisson equation reads
\begin{align}
  \lapl \psi(\vc\theta) = 2 \left[\sigma (\vc\theta) - \msigma\right] =:
  2\kappa (\vc\theta) \rtext{.}
\label{eq:poisson sphere gen}
\end{align}
We learn that not $\sigma$ itself is the source of the field but the
difference of $\sigma$ and the mean surface mass density
\begin{align}
  \msigma=\frac{1}{4\upi} \oiint\diff\Omega(\vc\theta) \,\sigma(\vc\theta)
  \rtext{.}
\end{align}

In Appendix~\ref{sec:phi2 from phi3 poiss} we show how the spherical lensing
potential can be written directly as integral over the Newtonian potential. It
is the three-dimensional Poisson equation that leads to the first term in
Eq.~\eqref{eq:poisson sphere gen}. The second term is related to the Newtonian
potential at the position of the observer. This tells us that the deviations
from the standard formalism can be related to local distortions of space-time.

\subsubsection{The magnification matrix}

The (inverse) magnification matrix is given by the Jacobian of the lens
mapping. On the tangential plane, the Jacobian of Eq.~\eqref{eq:lenseq} leads
directly to the Hessian of the potential, see Eq.~\eqref{eq:magn eq}.  The
situation is more complicated on the curved sphere, where finite displacements
cannot be treated as vectors.  In Appendix~\ref{sec:mag sph} we calculate the
magnification matrix for arbitrary deflection functions
$\vc\alpha(\vc\theta)$, including large deflection angles.  In terms of this
section, where the deflection angle $\vc\alpha$ is the gradient of a
potential $\psi$, the exact magnification matrix from Eq.~\eqref{eq:magn sph}
can be written in coordinates $(\parallel,\perp)$ parallel and perpendicular
to the negative deflection angle $-\vc\alpha$, as
\begin{align}
  \mat M^{-1} =
\begin{pmatrix}
  1-\psi_{\parallel;\parallel} && -\psi_{\parallel;\perp} \\[1.5ex]
  -\psi_{\perp;\parallel} \dfrac{\sin\alpha}{\alpha} && \cos\alpha -
  \psi_{\perp;\perp} \dfrac{\sin\alpha}{\alpha}
\end{pmatrix}
\rtext{.}
\label{eq:M psi alpha}
\end{align}
The lower indices of $\psi$ denote the second-order (covariant) derivatives
with respect to the coordinate axes.  The corrections in $\cos\alpha$ and
$\sin\alpha/\alpha$ are of second order in the deflection $\alpha$. These
terms can be set to unity if the magnification matrix is needed only to first
order. The result is then equivalent to the planar formalism in
Eqs.~(\ref{eq:hesse}--\ref{eq:magn eq}), but with the modified spherical
potential.

For a point mass at a distance $\theta$, the Hessian of $\psi$ is diagonal
with\footnote{
This can be derived from the covariant derivatives in $(\theta,\phi)$
coordinates, $T_{\theta;\theta}=T_{\theta,\theta}$,
$T_{\phi;\phi}=T_{\phi,\phi}+\sin\theta\cos\theta\, T_\theta$,
$T_{\theta;\phi}=T_{\phi;\theta}=T_{\theta,\phi}-\cot\theta\, T_\phi$, together
with the scaling transformation $\upartial_\parallel=\upartial_\theta$,
$\upartial_\perp=\upartial_\phi/\sin\theta$.}
\begin{align}
  \psi_{\parallel;\parallel} = \psi_{\theta,\theta}
\rtext{,} \qquad
\psi_{\perp;\perp} = \psi_{\theta} \cot\theta
\rtext{,}
\end{align}
which leads to
\begin{align}
  \mat M^{-1} =
\begin{pmatrix}
  1 + \dfrac{m}{4} \sin^{-2} \dfrac{\theta}{2} && 0
  \\[1.5ex]
  0 && \cos\alpha - \cot\theta \sin\alpha
\end{pmatrix}
\rtext{.}
\end{align}
This matrix corresponds exactly to Eqs.~(\ref{eq:mag tang}--\ref{eq:mag
  rad2}). Neglecting the curvature terms in $\mat M^{-1}$ corresponds to the
linear approximation in Eq.~(\ref{eq:mt lin}--\ref{eq:m m2 new}), where only
the term linear in $\alpha$ was included for the tangential magnification.

\subsubsection{Magnification theorem}

We have now, to first order, derived the magnification matrix as the Hessian
matrix of a potential, analogously to the standard formalism. We can thus
formally follow the proof for the magnification theorem as described in
Sec.~\ref{sec:theorem} step by step. However, the assumption of positive
(effective) $\kappa$ is no longer true, see Eq.~\eqref{eq:poisson sphere gen},
so that we cannot infer magnifications greater than unity from the signs of
the eigenvalues. Even though the density (even in comparison with the
reference situation) is still strictly non-negative, the convergence is not,
which invalidates one of the assumptions for the proof of the theorem.

Recall that, in order to define a firm foundation for this reasoning, we
refered to non-relativistic situations with unperturbed geometry in which the
deflection is caused either by refraction or by Newtonian deflection.
If we now return to the relativistic scenario but define the infinitely large
source sphere in such a way that the area is the same for the lensed and
unlensed situation, we come to exactly the same conclusion for that case.

\section{Sources at finite distances}
\label{sec:finite ds}

In the following we allow for sources at finite distances. In this situation,
the metric on the source sphere \emph{is} modified by the lens, so that the
previously used approach is not possible anymore, but assumptions about the
unlensed reference situation have to be made.  We do this by fixing the
coordinate distances, which would also be the appropriate approach for the
model of refraction or Newtonian deflection in an Euclidean metric.
Modifications of the results for different conventions (like fixed metric
distances) will be discussed later.

Main aim of this section is the analysis of lens deflection and the potential
theory on the sphere as opposed to the planar theory. For completeness this
has to be generalised to finite distances.  Most of the discussion of the
magnification theorem, on the other hand, is based on the case of sources at
infinity, because some arbitrariness in defining a reference situation can be
avoided then.

\begin{figure}
  \centering
  \includegraphics[width=0.475\textwidth]{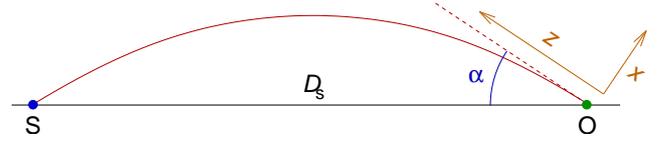}
\caption{The apparent deflection angle $\alpha$ is defined as the difference
  between the apparent and true source position. Changes of the direction of
  the light path will therefore affect $\alpha$ scaled with $\Dds(z)/\Ds$.}
\label{fig:geom3}
\end{figure}

\subsection{Deflection angle}

If the source is at a finite distance $\Ds$ from the observer, the simple
geometry in Fig.~\ref{fig:geom} has to be modified.  If we want to keep the
form of the lens equation \eqref{eq:lenseq}, the apparent deflection
angle $\vc\alpha$ is no longer defined as the angle between the incoming and
outgoing light ray, but as the difference between apparent and true source
position (Fig.~\ref{fig:geom3}). In the thin-lens approximation, this leads to
the well-known correction factor of $\Dds/\Ds$ to transform the true to the
apparent deflection angle.  As mentioned earlier, far from the optical axis
not even the point-mass lens can be considered as being thin, which means that
$\Dds$ is not defined globally.  If we instead consider how a local change of
$\dot{x}$ in Eqs.~\eqref{eq:alpha first} and following affects the difference
between true and apparent source position, we find that the integrands have
to be scaled with the \emph{local} $\Dds(z)/\Ds=(\Ds-z)/\Ds$. In addition, the
integration has to end at $z=\Ds$. With the same approach as before
($\lambda=z$), we find for the deflection angle of a point-mass lens
\begin{align}
  \alpha(\theta) &= -\int_0^{\Ds} \diff z\, \frac{\Ds-z}{\Ds}\,\ddot{x}
  (x=0,z)
\label{eq:alpha ds}
\\
&= -\frac{2GM}{c^2}\, x_0 \int_0^{\Ds} \diff z\, \frac{\Ds-z}{\Ds}\,\Bigl[
x_0^2 + (z-z_0)^2 \Bigr]^{-3/2} \\
&=-\frac{2GM}{c^2}\, \frac{1}{x_0\Ds} \left[
  \frac{(\Ds-z_0)^2+x_0^2}{\sqrtsmash{x_0^2+(\Ds-z_0)^2}} +
  \frac{z_0(\Ds-z_0)-x_0^2}{\sqrtsmash{x_0^2+z_0^2}}
\right] \\
&=\frac{2GM}{c^2}\, \frac{\Dds(\theta)+\Ds\cos\theta-\Dd}{\Dd\Ds\sin\theta}
\label{eq:alpha pm ds exact}
\rtext{,}
\end{align}
with the coordinate distance between source and mass
\begin{align}
  \Dds(\theta) = \sqrt{\Ds^2+\Dd^2-2\Dd\Ds\cos\theta} \rtext{.}
\end{align}
In the limit of small $\theta$ this reduces to\footnote{The Heaviside function
  is defined as $\heavi(x)=1$ for $x>0$ and $\heavi(x)=0$ otherwise.}
\begin{align}
  \alpha = \frac{4GM}{c^2} \left[ \frac{\Ds-\Dd}{\Dd\Ds}\,
    \frac{\heavi(\Ds-\Dd)}{\theta} + \ord{\theta} \right] \rtext{.}
\end{align}
For a source \emph{behind} the lens, we recover the classical limiting case
with the appropriate scaling of the apparent deflection angle. If the lens is
further away than the source (which we may call \emph{background lensing}),
the singularity at $\theta=0$ vanishes and the deflection generally becomes
much weaker, just as expected. In this case the maximum effect will be reached
at some finite $\theta$.

Note that in contrast to the standard expressions, Eq.~\eqref{eq:alpha pm ds
  exact} is valid for all combinations of distances. Even $\Ds\approx\Dd$ is
possible, including the limit of \emph{self-lensing} with $\Ds=\Dd$.

\begin{figure}
  \centering
  \includegraphics[width=0.47\textwidth]{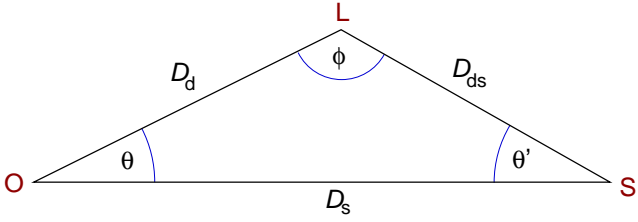}
\caption{Definition of the distance parameters $\Dd$, $\Ds$, $\Dds$ and the
  angles $\theta$, $\theta'$, and $\phi$ for a compact lens at {\sl L}.
  Observer and source are denoted by {\sl O} and {\sl S}. Recall that, in
  concordance with our general philosophy, the lengths and angles are in our
  convention defined in an Euclidean metric with the same coordinates as in
  the general metric of Eq.~\eqref{eq:weak field metric}. (See caption of Fig.~\ref{fig:geom}.)  }
\label{fig:dsfin}
\end{figure}

\subsection{Magnification}

Later we will compare magnification with amplification, which corresponds to
an exchange of observer and source, i.e.\ $\theta\leftrightarrow\theta'$ and
$\Dd\leftrightarrow\Dds$ (see Fig.~\ref{fig:dsfin}).  For this purpose, we
formulate the magnification in a form that explicitly shows the symmetries.  We
use the trigonometric relations
\begin{gather}
  \Ds^2 = \Dd^2+\Dds^2-2\Dd\Dds\cos\phi
  \rtext{,} \\
  \rlap{$\Ds \sin\theta$}\phantom{\Ds\cos\theta} = \Dds \sin\phi \rtext{,}
  \\
  \Ds \cos\theta = \Dd-\Dds \cos\phi \rtext{,}
\end{gather}
to write the tangential magnification $\mt$ following Eq.~\eqref{eq:mt lin}
for the deflection angle in Eq.~\eqref{eq:alpha pm ds exact} as
\begin{align}
  \mt^{-1} &= 1+\frac{GM}{c^2} \left( \frac{1}{\Dd} - \frac{1}{\Dds}
    -\frac{\Dd+\Dds}{\Dd\Dds} \tan^2\frac{\phi}{2} \right)
\label{eq:pm ds tang}
\rtext{.}
\end{align}
The radial magnification can be determined from the derivative of the
deflection angle, see Eq.~\eqref{eq:mag rad},
\begin{align}
  \mr^{-1} &= 1+\frac{GM}{c^2} \left( \frac{1}{\Dd} -\frac{1}{\Dds}
    +\frac{\Dd+\Dds}{\Dd\Dds} \tan^2\frac{\phi}{2}\right)
\label{eq:pm ds rad}
\rtext{.}
\end{align}
To first order in $\mat M$, the scalar magnification is given by
\begin{align}
  \mu^{-1} \eqlin 1+\frac{2GM}{c^2}\left( \frac{1}{\Dd} - \frac{1}{\Dds}\right)
\label{eq:omega pm ds mu}
\rtext{.}
\end{align}

Note that for these calculations we assumed that the coordinates of source and
observer are the same in the lensed and unlensed situation. For finite $\Ds$,
the lens will modify the geometry of the source sphere, which makes the
definition of the unlensed reference geometry somewhat arbitrary.  While
coordinates can be fixed for point-like sources, this is not appropriate for
the components of extended sources.  For those, we should keep their
\emph{physical size} constant to allow for a meaningful
comparison\footnote{Note that \citet{avni88}, who followed light rays from the
  source to the observer, took the changed geometry into account on the side
  of the source but not on the side of the observer. Even though they work in
  Schwarzschild coordinates, where the surface of a sphere with a constant
  radial coordinate $r$ is always $4\upi\, r^2$, this should not be neglected,
  since the comparison situation is a sphere around the source and not around
  the lens.}.

From the metric in Eq.~\eqref{eq:weak field metric} we infer that the ratio of
area to solid angle (on the source sphere) changes isotropically with the
potential $\Psis$ at the source position:
\begin{align}
  \diff A\sub s = \left( 1-\frac{2\Psis}{c^2} \right) \Ds^2\, \diff\Os
\end{align}
Using this we can define the true \emph{area} magnification $\mu_{(A)}$ as a
function of the \emph{solid angle} magnification $\mu_{(\Omega)}$, which we
called $\mu$ before:
\begin{align}
  \mu_{(\Omega)}^{-1} &= \frac{\diff\Os}{\diff\Omega} & \mu_{(A)}^{-1} &=
  \frac{1}{\Ds^2} \frac{\diff A\sub s}{\diff\Omega} = \left(
    1-\frac{2\Psis}{c^2} \right) \mu_{(\Omega)}^{-1}
\end{align}
The radial and tangential magnifications $\mr$ and $\mt$ both have to be
corrected with the same factor, which is the square root of that for the
scalar magnification, so that
\begin{align}
  \frac{{\mr}_{(A)}^{-1}}{{\mr}_{(\Omega)}^{-1}} =
  \frac{{\mt}_{(A)}^{-1}}{{\mt}_{(\Omega)}^{-1}} = 1-\frac{\Psis}{c^2} =
  1+\frac{GM}{c^2}\frac{1}{\Dds}
\rtext{.}
\end{align}
We now apply this correction to the magnifications from Eqs.~(\ref{eq:pm ds
  tang}--\ref{eq:pm ds rad}):
\begin{align}
  {\mt}_{(A)}^{-1} &= 1+\frac{GM}{c^2} \left( \frac{1}{\Dd}
    -\frac{\Dd+\Dds}{\Dd\Dds} \tan^2\frac{\phi}{2} \right)
\label{eq:A pm ds tang}
\\
{\mr}_{(A)}^{-1} &= 1+\frac{GM}{c^2} \left( \frac{1}{\Dd}
  +\frac{\Dd+\Dds}{\Dd\Dds} \tan^2\frac{\phi}{2} \right)
\label{eq:A pm ds rad}
\end{align}
The first-order scalar magnification is
\begin{align}
  \mu_{(A)}^{-1} \eqlin 1+\frac{2GM}{c^2} \frac{1}{\Dd}
\label{eq:A pm ds mu}
\rtext{.}
\end{align}

All these magnifications refer to an unlensed situation
with the same \emph{coordinate} distance to the source. As alternatives we
discuss fixed \emph{affine} and \emph{metric} distances in
Appendix~\ref{sec:alt distances}.

\subsection{Potential and Poisson equation}
\label{sec:poisson ds}

The potential can be derived by integrating the point-mass deflection angle in
Eq.~\eqref{eq:alpha pm ds exact}, or alternatively by integrating over the
appropriately scaled three-dimensional potential $\Psi$, as shown in
Appendix~\ref{sec:phi2 from phi3 pm ds} with the result in
Eq.~\eqref{eq:potential pm ds}.

In Appendix~\ref{sec:phi2 from phi3 poiss ds} we derive the Poisson equation
by integrating over the three-dimensional potential derivatives, and find the
relation
\begin{align}
  \lapl \psi(\vc\theta) = 2 \sigma (\vc\theta) + 2\frac{ \Psio -
    \Psis(\vc\theta)}{c^2}
\label{eq:poisson sigma ds}
\rtext{,}
\end{align}
where $\Psio$ is the Newtonian potential at the observer. The projected
surface mass density is now defined as
\begin{align}
  \sigma(\vc\theta) = \frac{4\upi G}{c^2} \int_0^{\Ds} \diff r\, r
  \frac{\Ds-r}{\Ds} \,\rho(\vc r)
\label{eq:sigma def2}
\rtext{.}
\end{align}
The weight function in Eq.~\eqref{eq:sigma def2} has the shape of a parabola
with maximum at $r=\Ds/2$ and zeros at $r=0$ and $r=\Ds$. However, since the
apparent size of a mass clump scales with $1/r^2$, the influence of a mass
element at $r$ is proportional to $1/r-1/\Ds$.

In classical lensing theory, the divergence of the deflection angle leads to
the local surface mass density. Far away from the optical axis but for sources
at infinity, this is changed in the way that we obtain in
Eq.~\eqref{eq:poisson sphere gen} the \emph{density contrast} relative to the
mean density, which results from the gravitational potential at the position
of the observer (see Appendix~\ref{sec:phi2 from phi3 poiss}).  Now we find in
Eq.~\eqref{eq:poisson sigma ds} for finite $\Ds$ that the divergence is also
affected by the potential at the source position (see Appendix~\ref{sec:phi2
  from phi3 poiss ds}). These two contributions describe the influence of
\emph{all} masses, while the surface mass density only includes masses inside
of the source sphere.  The projected surface mass density $\sigma$ vanishes
completely for $\Ds\le\Dd$, i.e.\ in the case where the source is closer to
the observer than the lens.

We recall from Sec.~\ref{sec:theorem} that the divergence of the deflection is
directly related to the first-order scalar magnification.  The same is true
here, but the corresponding magnification now refers to \emph{solid angles}
instead of area.  When writing Eq.~\eqref{eq:poisson sigma ds} for the
point-mass, we find that the potential terms lead directly to the correction
terms in Eq.~\eqref{eq:omega pm ds mu} if we follow the same recipe as in
Eq.~\eqref{eq:mu lin poisson}.

As consistency check we should test if the integral over Eq.~\eqref{eq:poisson
  sigma ds} vanishes, as required by the compact geometry of the sphere.  We
know from Gauss' theorem that the potential of a homogeneous spherical shell
is equivalent to that of a point-mass for regions outside of the shell, and
constant inside. We can turn the argument around, to learn that the potential
averaged over the surface of any sphere is the same as that in the centre of
the sphere, if masses are located only \emph{outside} of the sphere. These
masses do not influence the integral of the difference $\Psio-\Psis$ in
Eq.~\eqref{eq:poisson sigma ds}.  For masses \emph{inside} of the sphere, on
the other hand, the averaged potential is independent of their location, so
that the total mass can be thought of as being in the centre:
\begin{alignat}{2}
  <\Psis> &= -G \oiint \diff \Omega \int_0^{\Ds} \diff r \, r^2
  \frac{\rho(\vc r)}{\Ds} &&+ C \\
  \Psio &= -G \oiint \diff \Omega\, \int_0^{\Ds} \diff r \, r^2
  \frac{\rho(\vc r)}{r} &&+ C
\end{alignat}
The contributions from masses outside of the sphere are denoted as $C$.  By
comparison with Eq.~\eqref{eq:sigma def2}, we find
\begin{align}
  <\Psio - \Psis> &= -c^2 <\sigma> \rtext{.}
\end{align}
As required, the average source density in Eq.~\eqref{eq:poisson sigma ds}
vanishes.

\smallskip Note that for $\Ds\gg\Dd$ all expression in this section (and in
Appendix~\ref{sec:phi2 from phi3 poiss ds}--\ref{sec:phi2 from phi3 pm ds})
converge to the ones for infinite $\Ds$ derived in the previous sections and
Appendix~\ref{sec:phi2 from phi3 poiss}. The distinction between $\mu_{(A)}$
and $\mu_{(\Omega)}$ is not necessary in this limit.

\section{Reciprocity and surface brightness}
\label{sec:recip}

The magnifications in Eqs.~(\ref{eq:A pm ds tang}--\ref{eq:A pm ds mu}) are
not invariant under the exchange of source and observer.  Writing
$\mu_{(A)}(\text{o},\text{s})=\mu_{(A)}$ for the magnification of the source
as seen by the observer and $\mu_{(A)}(\text{s},\text{o})$ for the
magnification of the observer as seen by the source, we find (by exchanging
$\Dd\leftrightarrow\Dds$) the following relation, which is correct up to
first-order terms in the tangential and radial magnifications.
\begin{align}
  \frac{\mu_{(A)}(\text{s},\text{o})}{\mu_{(A)}(\text{o},\text{s})} = 1 +
  \frac{2GM}{c^2}\left( \frac{1}{\Dd}-\frac{1}{\Dds} \right)
\label{eq:recip1}
\end{align}
What does this mean physically? The \emph{magnification}
$\mu_{(A)}(\text{o},\text{s})$ defines the scaling of the apparent size of a
magnified source as seen by the observer. The reciprocal
$\mu_{(A)}(\text{s},\text{o})$, on the other hand, is inversely proportional
to the area in the observer plane that is spanned by a certain light bundle;
that means it defines the \emph{amplification} in terms of the number of
photons received from the source by the observer per detector area element.

The ratio of amplification to magnification in Eq.~\eqref{eq:recip1} provides
the gravitational change of `surface brightness', measured as photon number
density per solid angle. In this sense, gravitational lensing does \emph{not}
conserve surface brightness.

This may seem surprising but is in perfect agreement with the
\emph{reciprocity theorem} derived by \citet{etherington33}\footnote{See also
  \citet{ellis07} and the more easily available republication of the original
  article as \citet{etherington07}.}.  If one defines in an arbitrary
space-time for two events $x$ and $y$, which are connected by a null-geodesic,
i.e.\ one can be seen by the other, the angular size distances $D_{xy}$ for
the distance of $y$ as seen by $x$ and vice versa, it can be shown that the
two are related by
\begin{align}
  \frac{D_{yx}}{D_{xy}} = \frac{1+z_y}{1+z_x} \rtext{,}
\label{eq:recip2}
\end{align}
where $z$ are the redshifts as measured by an arbitrary observer.

In our situation, the lens magnifications act as corrections to obtain the
effective angular size distances from $\Ds$:
\begin{align}
  \frac{1}{D\sub{os}^2}&=\frac{\mu_{(A)}(\text{o},\text{s})}{\Ds^2} &
  \frac{1}{D\sub{so}^2}&=\frac{\mu_{(A)}(\text{s},\text{o})}{\Ds^2}
\label{eq:recip3}
\end{align}
The gravitational redshifts produced by the field of the lens can be derived
from the metric, Eq.~\eqref{eq:weak field metric}, as
\begin{align}
  1+z &= \left.\frac{c\,\diff t}{\diff s}\right|_{\diff\vc r=0} =
  1-\frac{\Psi}{c^2} = 1 + \frac{GM}{c^2} \frac{1}{D}
\label{eq:z}
\end{align}
for a point mass at a distance $D$.

We can confirm Eq.~\eqref{eq:recip1} by taking the ratio of
Eqs.~\eqref{eq:recip3} and inserting Eq.~\eqref{eq:recip2} and the redshifts
from Eq.~\eqref{eq:z}:
\begin{align}
  \frac{\mu_{(A)}(\text{s},\text{o})}{\mu_{(A)}(\text{o},\text{s})} &=
  \frac{D\sub{os}^2}{D\sub{so}^2} = \left(\frac{1+z\sub o}{1+z\sub s}\right)^2
  = \left(\frac{1+\dfrac{GM}{c^2} \dfrac{1}{\Dd}}{
      1+\dfrac{GM}{c^2} \dfrac{1}{\Dds}}\right)^2 \\
  &= 1 + \frac{2GM}{c^2} \left( \frac{1}{\Dd} - \frac{1}{\Dds} \right)
\end{align}

So far, we have discussed the surface brightness defined as \emph{photon
  number} per solid angle. If we want to determine it in terms of \emph{energy
  flux density} $F$, we have to take into account that both the energy of the
individual photons and the arrival rate at the observer are affected by the
redshift. This provides another factor of $(1+z\sub o)^2/(1+z\sub s)^2$, so
that the surface brightness in terms of intensity scales with
\begin{align}
  \frac{F\sub{obs}}{F_0} = \left(\frac{1+z\sub o}{1+z\sub s}\right)^4 = 1 +
  \frac{4GM}{c^2} \left( \frac{1}{\Dd} - \frac{1}{\Dds} \right) \rtext{.}
\end{align}
The scaling of the surface brightness with $(1+z)^{-4}$ is, of course, a
well-known fact in cosmology. Here we have exactly the same effect, but caused
by the metric perturbations of a gravitational lens at finite distance.  In
any case, we should keep in mind that the effect is extremely small, typically
of the order $m$ or the square of the Einstein radius. For typical
galaxy-scale lenses with Einstein radii of the order $\sim 1''$ this
corresponds to $\sim 10^{-11}\mbox{--}10^{-10}$.

\section{Light travel time}
\label{sec:light time}

We separate the light travel time into three parts: The undisturbed travel
time $\Ds/c$, and the geometrical and potential parts of the time-delay, $\tg$
and $\tp$, respectively.  The potential part is easy to calculate. We start
with the metric from Eq.~\eqref{eq:weak field metric} and write the
(coordinate) time interval for a null-curve along the $z$-axis as
\begin{align}
  \diff t = \left(1-\frac{2\Psi}{c^2}\right) \frac{\diff z}{c} \rtext{.}
\end{align}
For a point-mass lens, the integral is basically the same as for the metric
distance shown in Appendix~\ref{sec:metric}:
\begin{align}
  \tp &= \int \diff t - \frac{\Ds}{c} = -\frac{2}{c^3} \int_0^{\Ds}\diff
  z\, \Psi(x=0,z) \\
  &= \frac{2GM}{c^3} \ln\frac{\Dds(\theta)+\Dd+\Ds}{\Dds(\theta)+\Dd-\Ds} +
  \text{const}
\end{align}
We observe that this is proportional to the deflection potential in
Eq.~\eqref{eq:psi pm infty} only in the limit $\Ds\to\infty$, where we have
$\tp=-\Dd\,\psi/c$. In the general case there is no proportionality to
Eq.~\eqref{eq:potential pm ds}.

In fact, we do not expect a one-to-one relation between $\tp$ and the
potential $\psi$ as in the case of thin lenses. The argument used in
Sec.~\ref{sec:planar basics} to explain why the deflection angle is
proportional to the gradient of the potential time-delay breaks down, because
the geometrical part of the time-delay does not keep its simple form of
Eq.~\eqref{eq:fermat}. It cannot be expressed in terms of image and source
position alone anymore, because the deflection is not restricted to a lens
plane (or sphere).  In order to determine the geometrical delay, we have to
know the full three-dimensional mass distribution and not just its projection
$\sigma$ or the projected deflection potential $\psi$.

Because it cannot be easily generalized for arbitrary mass distributions, we
do not present the calculation of $\tg$ for a point-mass lens.

\begin{figure*}
  \centering
\ifpdf%
\includegraphics[width=0.485\textwidth]{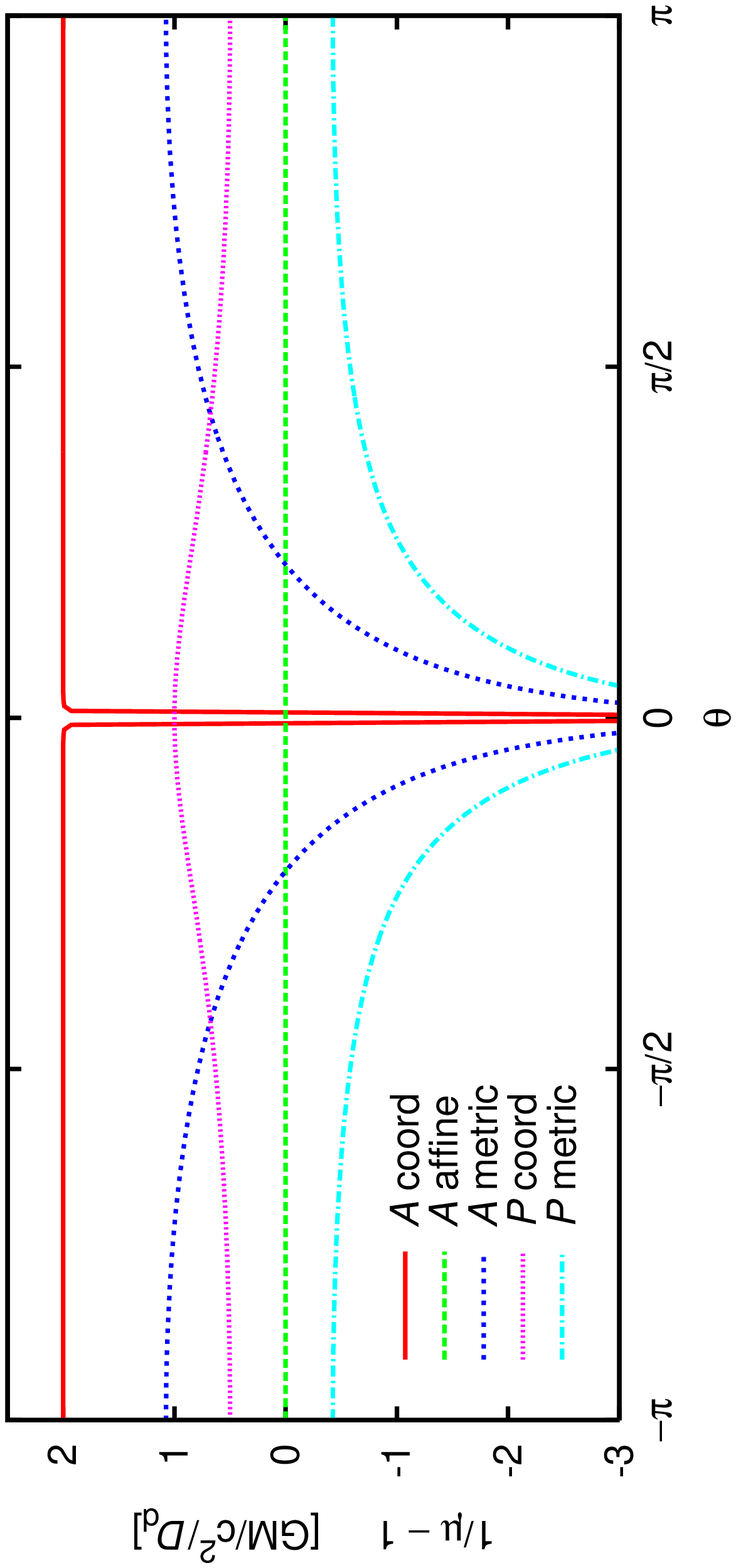}%
\else%
\includegraphics[angle=-90,width=0.485\textwidth]{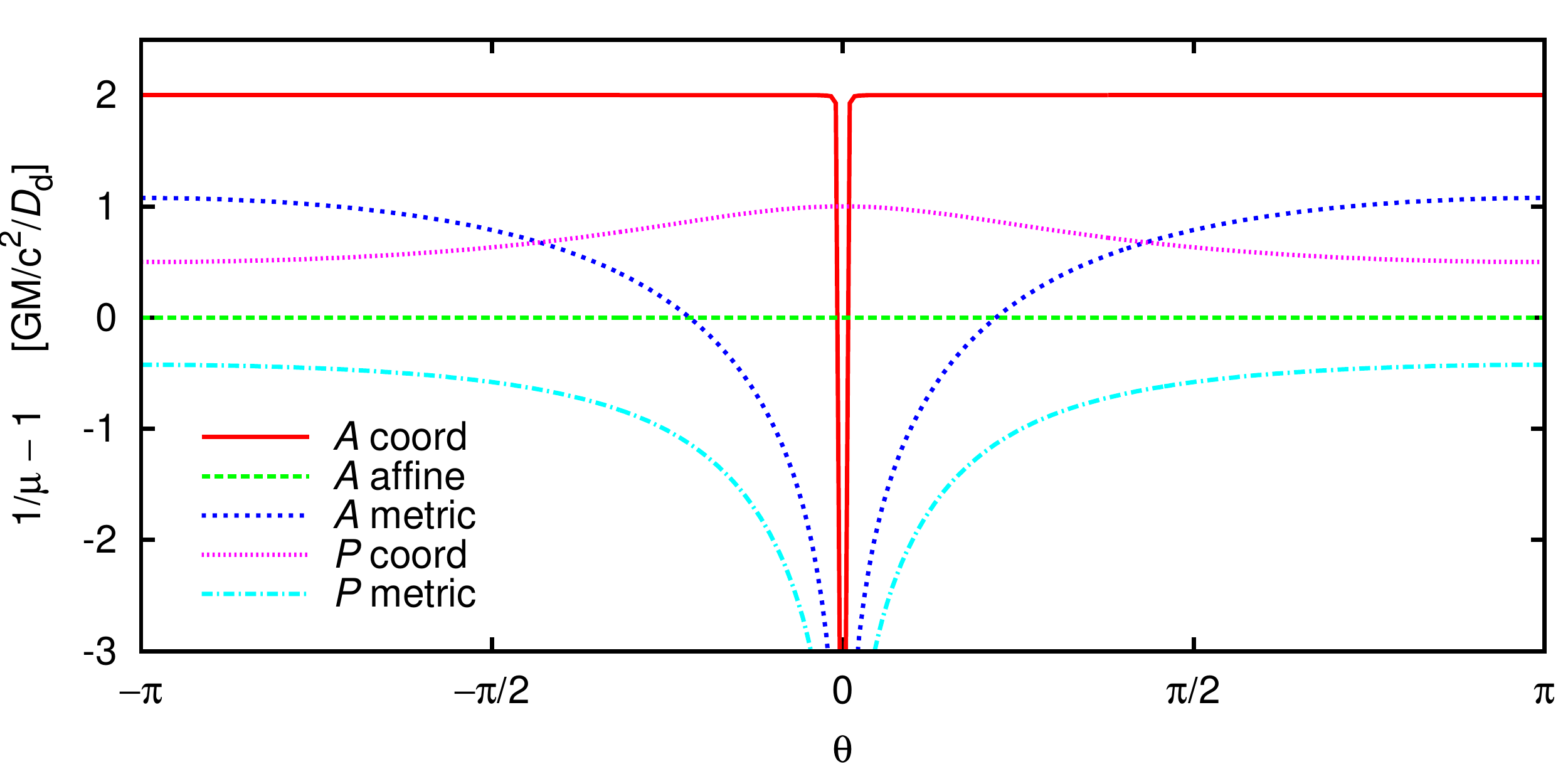}%
\fi%
\hspace{0.02\textwidth}%
\ifpdf
\includegraphics[width=0.485\textwidth]{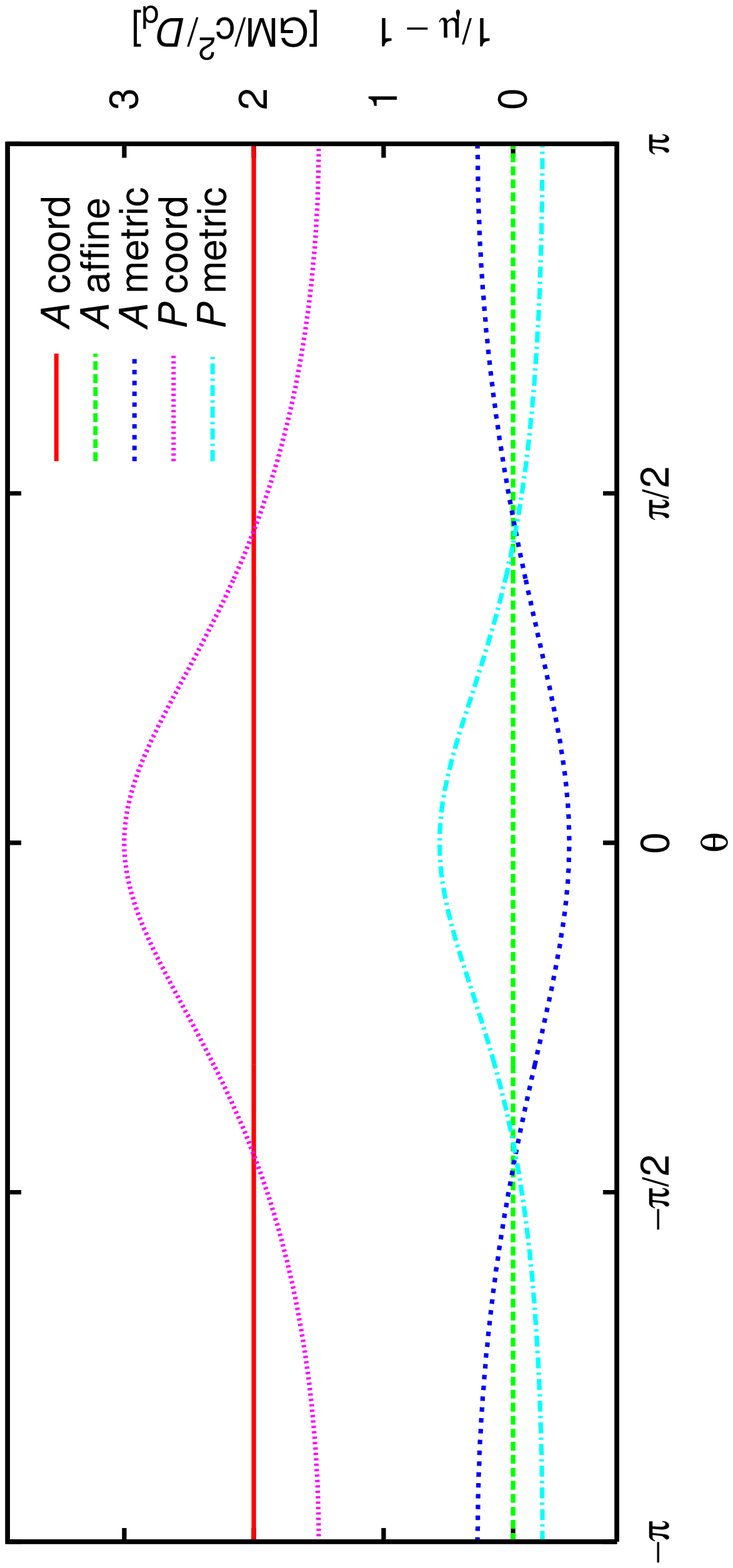}%
\else%
\includegraphics[angle=-90,width=0.485\textwidth]{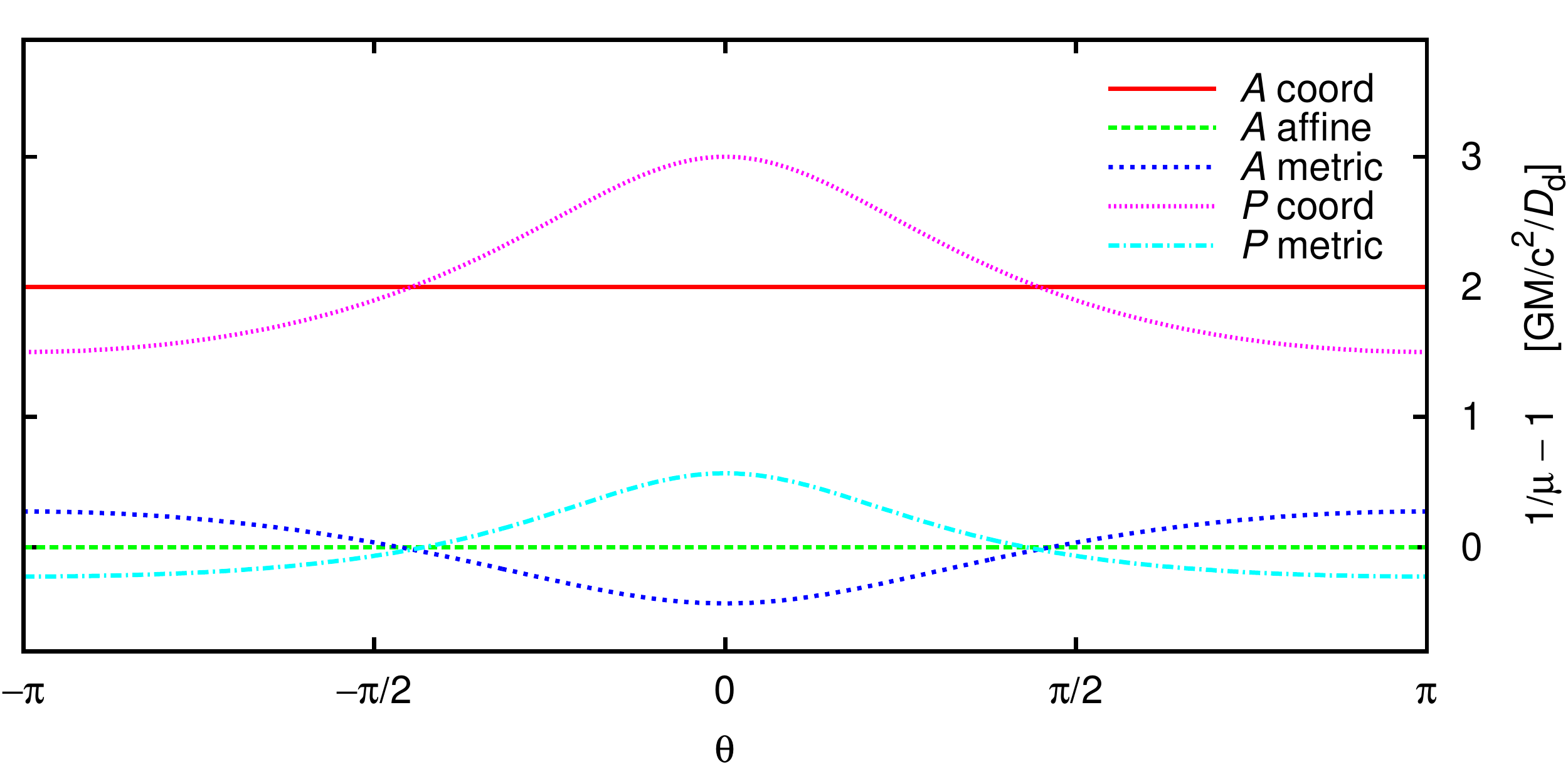}%
\fi%
\caption{Plots of several first-order magnifications defined in different
  ways for a point-mass lens. Labels $A$ refer to \emph{area magnification},
  $P$ to \emph{photon number amplification.} Different distances are held
  constant for the comparison with the unlensed situation: \emph{coordinate}
  distance, \emph{affine} distance and \emph{metric} distance. For `$A$ coord',
  we added the second-order magnification calculated for $GM/\Dd/c^2=10^{-8}$,
  which has no noticeable effect for $\Ds<\Dd$.  left: $\Ds=3\Dd$, right:
  $\Ds=\Dd/3$ }
\label{fig:1st order}
\end{figure*}

\section{Summary}
\label{sec:summary}

We discuss the classical magnification paradox in gravitational lensing in
order to better understand if and how a magnification greater than unity
everywhere can be consistent with global photon number conservation.  It is of
fundamental importance to allow for large angles relative to the optical axis.
Therefore we develop a formalism of lens and source \emph{spheres} instead of
\emph{planes}.  In this situation, even a point-mass lens cannot be regarded
as a \emph{thin} lens anymore. The `thickness' of a lens is not defined by the
extent of the mass distribution but by the extent of the deflecting potential
and its derivatives. The thickness of a point-mass lens is thus closely
related to the impact parameter and can be of the same order of magnitude as
the distances involved.

To avoid ambiguities, we start by considering a source at infinity, so that
the geometry on the source sphere is not changed by the action of the lens,
and the area of the source sphere is not modified. This makes a comparison
with the unlensed situation meaningful. We calculate magnifications by mapping
solid angle elements from the lens sphere to the source sphere. Because we do
not include horizons around black holes, which are corrections at higher
order, the total solid angle of the whole sky must be $4\upi$ with and without
lensing. Lensing defines a one-to-many mapping from the source sphere to the
lens sphere. We see the background sky in all directions, no parts of the
background are hidden, but some parts may be multiply imaged.  With this
argument, we conclude that the total magnification cannot be larger than unity
for all positions on the source sphere, in contradiction to the magnification
theorem.  This thought is supported by a simple geometric argument. The
density of deflected lines of sight on the source sphere (or equivalently
light rays on an observer's sphere around the source) can be increased close
to the optical axis, because a lens with positive mass has a focusing
tendency. In the approximation of the tangential plane, this can be true
everywhere, since lines of sight (and solid angle elements) can be borrowed
from infinity.  Once we consider the complete sphere, this is no longer
possible. Solid angle elements that are moved towards the optical axis must be
\emph{re-}moved from other parts of the sphere. The magnification cannot be
larger than one everywhere.  For this argument it is essential to work with
solid angles on a source sphere \emph{at infinity.}  With a sphere at finite
distance, the area density of projected lines of sight can well increase
everywhere, simply by making the area of the sphere smaller. Masses modify the
geometry, so that mean magnifications above one are no longer paradoxical.
The further the source sphere moves away, the weaker these distortions get.
For a finite mass, they decrease without limits and can finally be neglected.

The non-relativistic pictures of refraction and Newtonian deflection in
unperturbed geometry support this view. In these scenarios a paradox cannot
be avoided if no corrections to the formalism are made.

With this motivation, we calculate the deflection angle for a point-mass lens
for arbitrary angles to the optical axis with source at infinity and derive
magnifications from that. We find an additional first-order term that lowers
the magnification, violates the magnification theorem at some point, and
assures total conservation of solid angle.  In order to understand which part
of the classical proof of the theorem becomes invalid once we go from
tangential planes to full spheres, we develop the potential theory for
arbitrary mass distributions on the sphere. As in the plane, we can define a
two-dimensional lensing potential that is a projection of the
three-dimensional Newtonian potential.  In the same way we define a projected
surface mass density on the sphere.  We find that the Poisson equation is
modified in an interesting way.  In the plane, the divergence of the
deflection is proportional to the surface mass density. This cannot be true on
the sphere, because flux of the deflection field cannot escape from the
sphere. The integrated divergence must vanish in order to obey Gauss' law.  We
do indeed find that the divergence is $2(\sigma-\msigma)$, the density
contrast relative to the averaged density $\msigma$. The field lines
corresponding to the deflection field `decay' with increasing distance to the
masses.  This fact provides a very clean and firm background for a correction
of the standard solution of the paradox, according to which the magnification
theorem only holds compared to a reference Universe in which the lens is
\emph{removed}. If we instead redistribute the mass of the lens smoothly to
create a reference situation, the effective density contrast can become
negative, so that validity of the theorem is not expected anymore. Using the
formalism on the sphere in a perturbed Minkowski metric, we directly arrive at
the result that it is the density contrast relative to the average which
creates deflection and magnification.  This is true even though we use the
Minkowski metric with the lens \emph{removed} as our reference situation.  We
learn that at least in the asymptotic Minkowski metric, it does not make a
difference if we redistribute the mass of the lens over the lens sphere for
comparison, or if we simply remove it. A globally constant surface mass
density does note lead to light deflection\footnote{This does not contradict
  \citet{weinberg76}, where differences between the two cases are discussed.
  In our work, we consider a single, isolated lens, which distorts the
  geometry only locally, while \citet{weinberg76} discusses a Universe filled
  with lenses. In the latter case, the mean density and the global geometry
  \emph{do} depend on the presence of the lenses.}.

In the spherical potential formulation, the magnification matrix can be
calculated from the second-order derivatives of the potential, just as on the
plane. This neglects higher-order terms, which are not relevant in this
context.  However, in the Appendix we calculate exact magnifications for
arbitrarily large deflection angles. For this the curvature of the sphere has
to be taken into account fully.  The first-order formalism is analogous to
standard planar lensing. Nevertheless, the magnification theorem becomes
invalid, because the effective convergence $\kappa=\sigma-\msigma$ is not
strictly positive. Positivity of $\kappa$ was an essential assumption for the
proof, so that we can no longer expect the theorem to hold on the sphere.

To be able to switch the role of source and observer, so that magnification
becomes amplification, we have to allow sources at finite distances. We do
this be defining a source sphere with fixed coordinate distance from the
observer, using the isotropic weak-field metric.  We calculate the deflection
angle of a point mass and derive the magnification matrix from that. Here we
have to take into account that the metric on the source sphere is now modified
by the gravitation of the lens. We therefore have to distinguish between solid
angle and corresponding metric area. For the first-order area magnification,
we find the same result as for the source at infinity.  As before, the Poisson
equation has additional terms to $2\sigma$ on the source side. These are given
by the difference of the Newtonian potential at the positions of observer and
source. In the limit of a source at infinity, this corresponds again to the
mean surface mass density.  The solid angle magnification can be derived from
the potential in the usual way, but has to be corrected by the potential at
the source position to convert it into area magnification.

Alternative reference situations are treated as small perturbations of this
scenario, see Fig.~\ref{fig:1st order}. For a fixed metric distance, we find
that the magnification theorem still fails in some regions. This is different
if we fix the \emph{affine distance}, in which case the magnification theorem
holds everywhere as a direct result of the \emph{focusing theorem}
\citep[e.g.][]{perlick04}. This is even true for sources at infinity. However,
the affine distance is a highly inappropriate reference for astronomical
applications, because it is strongly influenced by local metric distortions at
the observer's position. While other, more practical, distance definitions,
like the metric distance, typically differ from the coordinate distance by the
order of the Schwarzschild radius of the lens, the affine distance is
\emph{scaled} by local perturbations. This is clearly not the reference
situation we would use in an astronomical scenario, because scaling the
distances changes the area even at infinity, even though the metric is not
affected at large distances from the lens.

An argument supporting this view is provided by the following
\emph{gedankenexperiment} (see Appendix~\ref{sec:small sphere}). The lens is a
thin spherical shell of radius $r_0$ around the observer, and the mass is
adjusted in such a way that $\sigma:=GM/(r_0c^2)$ is constant. If $\sigma$ is
sufficiently small, the fields are weak and all the approximations are valid.
For symmetry reasons, light rays reaching the observer will not be deflected
in this situation. This is consistent with our formalism and leads to a
magnification in solid angle of exactly unity in all directions.  In order to
come as close as possible to an unperturbed metric for the Universe, we now
make the radius $r_0$ arbitrarily small (scaling the mass $M$ appropriately).
In this limit, the metric perturbations are confined to an infinitesimal
region around the observer, and even there they are small.  Nevertheless, the
focusing theorem demands a magnification of $(1-2\sigma)^{-1}$ in all
directions, where the reference situation is defined to have an unchanged
affine distance to the source.  We can now imagine a practical experiment in
which we remove the lensing mass shell by shifting it slightly (but $\gg
r_0$). This corresponds to the `unlensed comparison situation' as the author
would define it in an astronomical context.  Since the mass of the lens is
infinitely small, the details of removing the lens are irrelevant for the
outcome, so that the reference situation is uniquely defined and corresponds
to our approach of keeping the coordinate (or metric) distance constant. The
experiment would find a magnification of exactly 1, in perfect agreement with
our reasoning.  The \emph{affine distance} for a fixed source, on the other
hand, would scale by a factor of $1/\sqrt{1-2\sigma}$ when the lens is put in
place.

An interesting variant of this gedankenexperiment uses values of $\sigma$ that
are not necessarily small, so that the weak-field approximation is no longer
valid. This has the invaluable advantage that the predicted \emph{diverging}
magnifications can seemingly be defined without deciding for a specific
reference situation (Schneider, priv.\ comm.).  However, we show in
Appendix~\ref{sec:small sphere} that magnifications in this case only diverge
for values of $\sigma$ for which the observer would be located within the
event horizon of the black hole which would then be formed by the lens. It is
clear that the magnifications in this definition do not mean that the observer
sees the source larger, but are purely related to the definition of the affine
distance, which scales with the relativistic time dilation at the observer.
We conclude that a fixed \emph{affine distance} does not define the unlensed
reference situation in a way most astronomers would like.

Comparing magnification and amplification, we find that the two are not equal
but are related by Etherington's reciprocity theorem. The ratio of
amplification and magnification defines the change of surface brightness in
terms of photon number, which is directly related to the gravitational
redshift of the source as seen by the observer. Concerning energy flux
density, the surface brightness is modified even stronger, because photon
energy and rate are also redshifted. These relations are completely equivalent
to the well-known relations in cosmology between angular size distance and
corrected and uncorrected luminosity distance.

Finally, we find that the deflection potential is no longer proportional to
the potential time-delay, as a result of different scaling of deflection and
Shapiro delay with distance. Even more disturbing for practical calculations,
the geometrical delay can no longer be expressed as a function of image and
source position alone. This well-known feature of thick lenses also affects
point-mass lenses in the case of large impact parameters.

It should be kept in mind that the modifications discussed in this article are
generally of extremely small magnitude, and it is difficult to imagine
situations in which they become directly measurable. Nevertheless, the
formalism does have practical applications, for example in the case of lenses
very close to or behind the sources, a case which is not covered by the
standard thin-lens approximation.

\section{Conclusions}
\label{sec:conclusions}

The total magnification excess obtained from a naive interpretation of the
magnification theorem is of the same order of magnitude as the perturbation of
the metric at the position of observer or source caused by the gravitation of
the lens. For a discussion of the apparent paradox, terms of
this order cannot be neglected.
This means that the lensing process can no longer be described in a tangential
plane, but full spheres have to be used for the lensed images and the unlensed
source.
Extending the planar formalism naturally leads to additional terms, which
compensate the magnification excess and lead to a mean magnification of
exactly 1. This is the case for an unlensed reference situation in which the
source sphere has the same area as the lensed source sphere.
The fact that total flux (or solid angle) is conserved in a formalism which is
based on this very assumption is clearly tautological. However, without the
additional spherical correction terms, this consistency could not have been
achieved.

It is a very satisfying result that generalising the formalism in order to
assure total flux conservation automatically avoids a violation of Gauss'
theorem for the deflection angle on the sphere, by modifying Poisson's
equation for the deflection potential. It is found that the source function
for the deflection is not the surface mass density itself, but the density
\emph{contrast} relative to the mean density. This result is entirely 
plausible, because a constant surface mass density can, for symmetry reasons,
not lead to a light deflection.  In our formalism, magnifications are defined
as directly being caused by light deflection and not by the indirect effect
of metric perturbations at the positions of source or observer. It is thus
clear that a constant surface mass density cannot lead to magnifications, and
that the magnification theorem does not hold on the complete sphere.

In contrast to this, a different comparison situation is explicitly defined
for the \emph{focusing theorem} \citep[e.g.][]{sef,perlick04}, in which the
affine distance is kept the same in the lensed and unlensed situation.
With this definition, the theorem is \emph{always} valid, even under much more
general circumstances. However, we argue that keeping the affine distance
constant is not the appropriate way to compare lensed and unlensed sources in
an astronomical context.

In addition we learn that at the level of accuracy needed for a discussion of
the magnification theorem(s), we also have to distinguish between
magnification and amplification, and that for the latter we have to decide if
we define it in terms of total photon number, photon flux, total energy or
energy flux. This distinction is well known in the context of cosmological
redshifts. As a result of these effects, surface brightness is generally
\emph{not} conserved.

\section*{Acknowledgments}

The author likes to thank Richard Porcas and Peter Schneider for
inspiring and controversial discussions about the subject.
This work was supported by the European Community's Sixth Framework Marie
Curie Research Training Network Programme, Contract No.\ MRTN-CT-2004-505183
``ANGLES'', and by the Emmy-Noether-Programme of the `Deutsche
Forschungsgemeinschaft', reference Wu\,588/1-1.

\newcommand{\mnras}{\mbox{MNRAS}}
\newcommand{\apj}{\mbox{ApJ}}
\newcommand{\apjl}{\mbox{ApJ}}
\newcommand{\aap}{\mbox{A\&A}}
\newcommand{\prd}{\mbox{Phys.~Rev.~D}}

\bibliographystyle{mn2e}
\bibliography{paper}

\appendix

\section{Spherical potential as integrated three-dimensional potential}

\subsection{Poisson equation for source at infinity}
\label{sec:phi2 from phi3 poiss}

We define a two-dimensional potential $\psix$ on the sphere as a certain
projection of the Newtonian potential $\Psi$,
\begin{align}
  \psix(\vc\theta) =\frac{2}{c^2} \int_0^\infty \diff r \, \frac{\Psi(\vc
    r)}{r} +\text{const}
\label{eq:phi2 from phi3}
\rtext{,}
\end{align}
where we have to expect problems with the convergence and may have to subtract
diverging terms that are independent of $\vc\theta$.  If we take the angular
gradient $\nablarhat=r\nablaperp$ of this potential, we recover the previous
equation \eqref{eq:alpha int Psi} for the deflection angle. We can therefore
omit the tilde and write $\psi$ instead.

Now we want to calculate the two-dimensional angular Laplacian of the
potential. We have $\laplrhat=r^2\laplperp$ and thus
\begin{align}
  \laplrhat\psi(\vc\theta) = \frac{2}{c^2} \int_0^\infty \diff r \, r\,
  \laplperp\Psi(\vc r) \rtext{.}
\end{align}
The integrand can be calculated by using the three-dimensional Laplacian
\begin{align}
  \lapliii\Psi = \laplperp\Psi +
  \frac{1}{r}\frac{\upartial^2(r\Psi)}{\upartial r^2}
\end{align}
and the three-dimensional Poisson equation
\begin{align}
  \lapliii\Psi (\vc r) = 4\upi G \, \rho(\vc r) \rtext{.}
\end{align}
This leads to
\begin{align}
  \laplrhat\psi(\vc\theta) &= \frac{2}{c^2} \left\{ 4\upi G \int_0^\infty \diff
    r \, r\, \rho(r) - \int_0^\infty \diff r \,
    \frac{\upartial^2\left[r\Psi(\vc r)\right]}{\upartial r^2}
  \right\} \\
  &= \frac{8\upi G}{c^2} \int_0^\infty \diff r \, r\, \rho(r) -
  \frac{2}{c^2}\left[\Psi(\vc r) + r \frac{\upartial \Psi(\vc r)}{\upartial
      r} \right]_0^\infty
\label{eq:phi2 from phi3 int}
\rtext{.}
\end{align}
We assume that the mass is localized and finite, so that the second term
vanishes at infinity. We are left with the potential $\Psio=\Psi(0)$ at the
position of the observer:
\begin{align}
  \laplrhat\psi(\vc\theta) = \frac{8\upi G}{c^2} \int_0^\infty \diff r \, r\,
  \rho(\vc r) + \frac{2}{c^2} \Psio
\label{eq:int r psi0}
\end{align}
We use
\begin{align}
  \Psio &= -G \iiint\diff^3 r \frac{\rho(\vc r)}{r} \\
  &= -G \int_0^\infty \diff r \, r\oiint \diff\Omega(\vc\theta)\,\rho(\vc r) \\
  &= -4\upi G \int_0^\infty \diff r \, r\,\mrho(r) \rtext{,}
\end{align}
with the mean density at radius $r$,
\begin{align}
  \mrho(r) = \frac{1}{4\upi} \oiint \diff\Omega(\vc\theta)\,\rho(\vc r)
  \rtext{.}
\end{align}
After inserting this into Eq.~\eqref{eq:int r psi0}, we find
\begin{align}
  \laplrhat\psi(\vc\theta) = \frac{8\upi G}{c^2} \int_0^\infty \diff r \, r\,
  \left[\rho(\vc r) -\mrho(r) \right] \rtext{.}
\end{align}
With the definition for the normalized surface mass density from
Eq.~\eqref{eq:sigma def1} and the equivalent for $\msigma$, we finally arrive
at the Poisson equation \eqref{eq:poisson sphere gen}:
\begin{align}
  \lapl \psi(\vc\theta) = 2 \left[\sigma (\vc\theta) - \msigma\right]
\end{align}

\subsection{Poisson equation for source at finite distance}
\label{sec:phi2 from phi3 poiss ds}

We can generalize the calculations from last section for finite $\Ds$. For
this purpose we have to stop the integration at $\Ds$ and radially scale the
potential in the same way as the deflection angle, compare Eq.~\eqref{eq:alpha
  ds} with \eqref{eq:alpha first}.  This leads to
\begin{align}
  \psi(\vc\theta) = \frac{2}{c^2} \int_0^{\Ds} \diff r \, \frac{\Ds-r}{r\,
    \Ds}\Psi(\vc r) \rtext{,}
\label{eq:psi2 from psi3 ds}
\end{align}
and after integrating in the same way as before, where we now in addition to
the potential at the observer's position $\Psio$ also have a contribution from
the potential at the source position $\Psis$, to
\begin{align}
  \lapl \psi(\vc\theta) = 2 \left[ \sigma (\vc\theta) + \frac{ \Psio - \Psis
      (\vc\theta)}{c^2} \right] \rtext{.}
\end{align}
The normalized surface mass density $\sigma$ is now defined according to
Eq.~\eqref{eq:sigma def2}. Please note that the additional potential terms can
no longer be calculated from the surface mass density alone, because they also
depend on masses outside of the source sphere, which do not contribute to
$\sigma$.

\subsection{Potential for point mass at finite distance}
\label{sec:phi2 from phi3 pm ds}

We start with Eq.~\eqref{eq:psi2 from psi3 ds} and insert the potential for a
point mass at the position defined in Eq.~\eqref{eq:x0z0}.
\begin{align}
  \psi(\theta) &= -\frac{2GM}{c^2} \int_0^{\Ds}\diff z \, \frac{\Ds-z}{z\,\Ds}
  \frac{1}{\sqrtsmash{x_0^2+(z-z_0)^2}}
\end{align}
This integral is divergent at the lower limit, but can be regularised, and
after some simplifications reduces to
\begin{multline}
  \psi(\theta) = \frac{2GM}{c^2} \, \Biggl[
  \frac{\ln \left[\Dd-\Ds\cos\theta+\Dds(\theta)\right]}{\Dd} \\
  + \frac{\ln
    \left[\Ds-\Dd\cos\theta+\Dds(\theta)\right]-\ln(1-\cos\theta)}{\Ds}
  \Biggr]
\label{eq:potential pm ds}
\rtext{.}
\end{multline}

\section{Exact magnification matrix on the sphere}
\label{sec:mag sph}

We want to calculate the magnification matrix for arbitrary functions of the
deflection angle. This includes large deflections and multi-plane lenses,
where the deflection angle can no longer be written as the gradient of a
potential. This generality is not necessary for the main part of this paper
but may serve as the basis for future work.

In the plane, the total displacement is not relevant, so that the
magnification matrix is determined exclusively by the first-order derivatives
of the deflection.  On the sphere, we have to take into account the curvature,
and the lens equation is no longer a vector equation.

To determine the source position $\Theta\sub s$ from the image position
$\Theta$, we have to move along a geodesic (or great circle) in the direction
of the negative deflection angle and follow this geodesic for a length
corresponding to the absolute deflection angle. The geodesic equation for
arbitrary coordinates is
\begin{align}
  \ddot x^\alpha + \Gamma^\alpha_{\mu\nu} \dot x^\mu \dot x^\nu = 0 \rtext{.}
\end{align}
The affine parameter $\lambda$ runs from 0 at $\Theta$ to 1 at $\Theta\sub s$.
Derivatives with respect to $\lambda$ are written as dots. In the following, we
write the deflection angle as $a^\mu$ (with $a^2 = a^\mu a_\mu$) to avoid
confusion with tensor indices. The boundary conditions are
\begin{align}
  x^\alpha (0) &= \Theta^\alpha \rtext{,} & \dot x^\alpha (0) &= -a^\alpha
  \rtext{,}
\label{eq:boundary x} &
x^\alpha (1) &= \Theta\sub s^\alpha \rtext{.}
\end{align}
For the magnification matrix, we have to consider additional geodesics
infinitely close to the reference geodesic. The equation for the difference
$\epsilon\xi^\alpha$, where $\epsilon$ is infinitely small, is the
differential equation for the geodesic deviation:
\begin{align}
  \frac{\Diff^2 \xi^\alpha}{\Diff\lambda^2} = \dot x^\beta \dot x^\mu \xi^\nu
  R^\alpha_{\mu\beta\nu}
\label{eq:geodevi cov}
\end{align}
The differential operator $\Diff$ denotes covariant derivatives. The curvature
tensor $R$ has a particularly simple form for two-dimensional manifolds. It
can be written in terms of the metric $g_{\mu\nu}$ as
\begin{align}
  R^\alpha_{\mu\beta\nu} &:=
  \Gamma^\alpha_{\mu\nu,\beta}-\Gamma^\alpha_{\mu\beta,\nu} +
  \Gamma^\alpha_{\rho\beta}\Gamma^\rho_{\mu\nu}-
  \Gamma^\alpha_{\rho\nu}\Gamma^\rho_{\mu\beta} \\
  &\phantom{:}= \frac{1}{K^2} \left( \delta^\alpha_\beta g_{\mu\nu} -
    \delta^\alpha_\nu g_{\mu\beta} \right) \rtext{.}
\label{eq:curv tensor}
\end{align}
The curvature radius $K$ is constant ($K=1$) on the sphere. The limit of the
tangential plane can be found as $K\to\infty$.  Eq.~\eqref{eq:geodevi cov} is
valid in any coordinate system. For our convenience we use the system defined
by the coordinates at $\Theta$, which is then parallel-transported along the
geodesic. In this way, the covariant derivatives become partial derivatives of
the components, and $\dot x^\alpha\equiv(a,0)$ as well as the curvature tensor
Eq.~\eqref{eq:curv tensor} have constant components. We use a local Cartesian
system (with locally vanishing Christoffel symbols) in which $\xi^\parallel$
is measured parallel to the negative deflection angle and $\xi^\perp$
orthogonal to this direction. This leads to
\begin{align}
  \ddot\xi^\parallel = 0 \rtext{,} \qquad \ddot\xi^\perp = -\omega^2 \xi^\perp
  \rtext{,} \qquad \omega:=\frac{a}{K} \rtext{.}
\label{eq:geodevi cov2}
\end{align}
With the starting condition
\begin{align}
  \dot\xi^\alpha = \frac{\Diff \xi^\alpha}{\Diff \lambda} = -\xi^\mu
  \frac{\Diff a^\alpha}{\Diff x^\mu}
\end{align}
from the derivative of Eq.~(\ref{eq:boundary x}b), we can easily solve the
differential equation \eqref{eq:geodevi cov2} for the two starting vectors
$(1,0)$ and $(0,1)$ and in this way write the transport equation from $\Theta$
to $\Theta\sub s$ for arbitrary vectors $\xi^\mu$ as
\begin{align}
  \vc\xi (1) = \mat M^{-1} \vc\xi (0) \rtext{.}
\end{align}
The inverse magnification matrix of this mapping in $(\parallel,\perp)$
coordinates reads
\begin{align}
  \mat M^{-1} =
\begin{pmatrix}
  1-a^\parallel_{;\parallel} && -a^\parallel_{;\perp} \\[1.5ex]
  -a^\perp_{;\parallel} \dfrac{\sin\omega}{\omega} && \cos\omega -
  a^\perp_{;\perp} \dfrac{\sin\omega}{\omega}
\end{pmatrix}
\rtext{,}
\label{eq:magn sph}
\end{align}
in terms of the derivatives (covariant or partial in these coordinates) of the
deflection function $a^\mu$. The magnification depends on the derivatives, but
also on the deflection angle itself, just as expected.  We notice that finite
deflection angles introduce \emph{rotation} even if the deflection field is
rotation-free ($a^\parallel_{;\perp}=a^\perp_{;\parallel}$). Furthermore does
the curvature of the sphere lead to a magnification of $1/\cos\omega$ in the
perpendicular direction even for (locally) constant deflection fields. In the
interpretation of this, one should keep in mind that a covariantly constant
deflection does \emph{not} correspond to a rigid rotation of the sphere, not
even locally.

We can decompose the inverse magnification matrix into a rotated convergence
and a shear part,
\begin{align}
  \mat A := \mat M^{-1} = (1-\kappa) \begin{pmatrix} \cos\varphi &
    -\sin\varphi \\ \sin\varphi & \hfill\cos\varphi
\end{pmatrix}  - \begin{pmatrix}
  \gamma_1 & \hfill\gamma_2 \\ \gamma_2 & -\gamma_1
\end{pmatrix}
\rtext{,}
\end{align}
where the parameters are determined by the following equations:
\begin{align}
  \tan\varphi &= \frac{A_{21}-A_{12}}{A_{11}+A_{22}} =
  \frac{a^\parallel_{,\perp}-a^\perp_{,\parallel} \sin\omega/\omega
  }{1+\cos\omega-a^\parallel_{,\parallel} - a^\perp_{,\perp}
    \sin\omega/\omega}
\label{eq:mag sph phi}
\\
1-\kappa &= \frac{\sign(A_{11}+A_{22})}{2}\sqrt{(A_{11}+A_{22})^2 +
  (A_{12}-A_{21})^2}
\label{eq:mag sph kappa}
\\
\gamma_1 &= -\frac{A_{11}-A_{22}}{2} = \frac{\cos\omega-1
  +a^\parallel_{,\parallel} - a^\perp_{,\perp} \sin\omega/\omega}{2}
\\
\gamma_2 &= -\frac{A_{12}+A_{21}}{2} = \frac{a^\parallel_{,\perp} +
  a^\perp_{,\parallel} \sin\omega/\omega}{2}
\end{align}
Note that the mapping is invariant under a sign change of $1-\kappa$ with a
simultaneous shift of $\upi$ in $\varphi$.  Equations \eqref{eq:mag sph phi}
and \eqref{eq:mag sph kappa} are consistent for $-\upi/2<\varphi<\upi/2$.

In the limit of small $\omega$ (corresponding to small deflection angles or
$K\to\infty$), the matrix reduces to the standard form
\begin{align}
  \mat M^{-1} =
\begin{pmatrix}
  1-a^\parallel_{;\parallel} & -a^\parallel_{;\perp} \\[1ex]
  -a^\perp_{;\parallel} & 1 - a^\perp_{;\perp}
\end{pmatrix}
\rtext{.}
\end{align}

\section{Alternative distance measures}
\label{sec:alt distances}

\subsection{Affine distance}
\label{sec:affine}

The so-called \emph{focusing theorem} \citep[e.g.][]{perlick04} is similar to
the magnification theorem but is valid for arbitrary space-times, including
strong fields and regions far from the optical axis. It says that an image
created before the first conjugate point (before parity inversion) is always
magnified with a factor $|\mu_\Lambda|\ge 1$ compared to an unlensed source
\emph{at the same affine distance}. This is different from our discussion,
because we kept the \emph{coordinate distance} fixed.

The affine distance is measured as the affine parameter of a null-geodesic
connecting two events, where the parameter is normalized to approach metric
distance very close to the observer. To compare this with our result, where
the magnification can be less than unity, we have to calculate the affine
distance of the lensed image.  We are only interested in first-order effects
in $\Psi$, which makes this task easy. Deviations of the geodesic from a
straight line in coordinates would be of second order and can be neglected. We
thus only need one spatial coordinate $x$ measured along the geodesic. We use
the metric from Eq.~\eqref{eq:weak field metric} to define a Lagrange function
for a geodesic, writing derivatives with respect to the affine parameter
$\lambda$ as dots,
\begin{align}
  \lagrange = \left( \frac{\diff s}{\diff\lambda} \right)^2 = \left(
    1+\frac{2\Psi}{c^2} \right) c^2\dot t^2 - \left( 1-\frac{2\Psi}{c^2}
  \right) \dot x^2 \rtext{.}
\end{align}
With the Euler-Lagrange equations, we derive to first order $\dot x =
\text{const}$, and with the normalization at the observer, find
\begin{align}
  \dot x = 1+ \frac{\Psio}{c^2} \rtext{,}
\end{align}
where $\Psio$ is the potential at the observers position. This means that the
affine distance $\Lambda$ is the coordinate distance scaled with the constant
$1/\dot x$. To redefine the magnification as a comparison with an unlensed
source at the same affine distance, we have to scale it with
$(\Lambda/\Ds)^2$. For a compact lens of mass $M$ this leads to the
first-order correction
\begin{align}
  \mu_\Lambda = \mu \left( 1- \frac{2\Psio}{c^2} \right) = \mu \left( 1+
    \frac{2GM}{c^2\Dd} \right) = \mu \left( 1+ \frac{m}{2} \right) \rtext{.}
\end{align}
For $\Ds\to\infty$ this cancels the $m/2$ term in Eq.~\eqref{eq:m m2 new},
which produced magnifications less than unity for large $\theta$.  The same is
true for the case with finite $\Ds$ in Eqs.~(\ref{eq:A pm ds tang}--\ref{eq:A
  pm ds mu}).  See Fig.~\ref{fig:1st order} for a graphical illustration.

\subsection{Metric distance}
\label{sec:metric}

Alternatively, we could for the comparison of lensed and unlensed situation
keep the \emph{metric distance} $L$ constant, which we define as measured with
a rigid ruler being at rest in the $(t,\vc r)$ coordinates of the weak-field
metric.  This can be integrated easily for a point mass at the standard
position of Eq.~\eqref{eq:x0z0}:
\begin{align}
  L &= \int \sqrt{-\diff s^2}\bigr|_{\diff t=0} = \int_0^{\Ds} \diff z\,
  \left[ 1-\frac{\Psi(x=0,z)}{c^2}\right]  \\
  &= \Ds + \frac{GM}{c^2} \int_0^{\Ds} \diff z\,
  \frac{1}{\sqrtsmash{x_0^2+(z-z_0)^2}}
  \\[0.5ex]
  &= \Ds + \frac{GM}{c^2}
  \ln\frac{\Ds-\Dd\cos\theta+\Dds(\theta)}{\Dd(1-\cos\theta)} \\
  &= \Ds + \frac{GM}{c^2} \ln
  \frac{\Dds(\theta)+\Dd+\Ds}{\Dds(\theta)+\Dd-\Ds} \rtext{.}
\end{align}
To correct the magnifications to be valid for fixed $L$, they have to be
multiplied with $(L/\Ds)^2$ in the same way as above. Results are shown in
Fig.~\ref{fig:1st order}.

\section{Lensing by a spherical shell}
\label{sec:small sphere}

A very instructive example is that of a thin spherical shell acting as a lens,
with the observer in the centre of the shell and the source sphere at
infinity. Due to the symmetry, radial 
light rays are not deflected, so that we do not expect magnification caused by
deflection. In the formalism of this paper, see Eq.~\eqref{eq:poisson sphere
  gen}, the constant surface mass density leads to a constant potential and no
deflection or magnification.
In the formalism of the focusing theorem, on the other hand, where the affine
distance is kept constant to compare the lensed and unlensed situation, we
still have focusing without deflection.
One might think that the case of infinite focusing would allow a discussion
without the arbitrariness of distance definitions, because different distances
can only \emph{rescale} the focusing.

In order to discuss this case, we have to leave the weak-field approximation
and use the full metric for a shell with total mass $M$, $\Rm=GM/c^2$ and
radius $r_0$ in Schwarzschild coordinates:
\begin{align}
\diff s^2 &= \begin{cases}
\displaystyle
\left( 1-\frac{2\Rm}{r_0} \right) \, \diff t^2 - \diff r^2 - r^2\,\diff\Omega^2 
& \text{for $r<r_0$} \\
\displaystyle
\left( 1-\frac{2\Rm}{r} \right) \, \diff t^2 -
\left(1-\frac{2\Rm}{r}\right)^{-1}\, \diff r^2 - r^2\,\diff\Omega^2 
& \text{for $r>r_0$}
\end{cases}
\end{align}
We are especially interested in the limiting case of undistorted global
geometry of the Universe, therefore we consider very small $r_0$ and $M$, but
keep $\Rm/r_0$ constant. Using Eq.~\eqref{eq:sigmac} it can be shown that
\emph{in the weak field limit} we have 
\begin{align}
\sigma &= \frac{\Rm}{r_0}
\rtext{,}
\end{align}
which we extend \emph{as a definition} of $\sigma$ for strong fields.
In the limit of small $\Rm$, the metric is distorted only in a small region
around the 
observer. The radial metric distance then corresponds to $r$. The affine
distance can be calculated with methods similar to those described in
Appdendix~\ref{sec:affine}, which results in
\begin{align}
\Lambda &= \frac{r}{\sqrt{1-2\sigma}}  \approx (1+\sigma)\,r 
\rtext{.}
\end{align}
The magnification defined for an unchanged affine distance scales with
\begin{align}
\mu_\Lambda &= \left(\frac{\Lambda}{r}\right)^2 = \frac{1}{1-2\sigma}
\rtext{,}
\end{align}
which obeys the focusing theorem $\mu_\Lambda> 1$.

The critical case $\sigma=1/2$ corresponds to a lens which forms a black hole,
which is clearly beyond the weak-field approximation. Even in more moderate
cases, we see that the ``focusing'' is purely a result of the local
measurement of the affine distance, which is affected by the metric
distortions.

\bsp

\end{document}